\documentclass[iop]{emulateapj}
\usepackage{natbib}
\usepackage{apjfonts}
\usepackage{epsfig}
\usepackage{longtable}
\usepackage{array}

\newcommand{\rhosfr}{\dot{\rho}_{\star}}

\newcommand{\GRBSFR}{\Psi}
\newcommand{\GRBSFRl}{\psi}
\newcommand{\SFR}{\mathrm{SFR}}
\newcommand{\Msun}{M_{\sun}}
\newcommand{\Swift}{Swift}
\newcommand{\za}{z_{a}}
\newcommand{\zb}{z_{b}}

\newcommand{\zlim}{z_{\mathrm{max}}}
\newcommand{\Liso}{L_{\mathrm{iso}}}
\newcommand{\Eiso}{E_{\mathrm{iso}}}
\newcommand{\Mcrit}{M_{\star,\mathrm{crit}}}
\newcommand{\Mstar}{M_{\star}}
\newcommand{\Phistar}{\Phi_{\star}}
\newcommand{\phistar}{\phi_{\star}}
\newcommand{\Zcrit}{12+\log[\mathrm{O}/\mathrm{H}]_{\mathrm{crit}}}
\newcommand{\Ngrb}{N_{\mathrm{GRB}}}
\newcommand{\fesc}{f_{\mathrm{esc}}}
\newcommand{\tnm}[1]{\tablenotemark{#1}}

\shorttitle{}
\shortauthors{Robertson \& Ellis}

\begin{document}

\title{Connecting the Gamma Ray Burst Rate and the Cosmic Star Formation History: Implications for Reionization and Galaxy Evolution}

\author{Brant E. Robertson\altaffilmark{1, 2, 3} and Richard S. Ellis\altaffilmark{1}}

\altaffiltext{1}{Astronomy Department, California Institute of Technology, 
MC 249-17, 1200 East California Boulevard, Pasadena, CA 91125}
\altaffiltext{2}{Steward Observatory, University of Arizona, 933 North Cherry Avenue, Tucson, AZ 85721}
\altaffiltext{3}{Hubble Fellow, brant@astro.caltech.edu}

\begin{abstract}

The contemporary discoveries of galaxies and gamma ray bursts (GRBs) at high redshift
have supplied the first direct information on star formation when the universe was only
a few hundred million years old.  The probable origin of long duration GRBs 
in the 
deaths of massive stars would link the universal GRB rate to the 
redshift-dependent star formation rate density, although exactly how is 
currently unknown.  As the most distant GRBs and star-forming galaxies probe
the reionization epoch, the potential rewards of understanding the redshift-dependent ratio
$\GRBSFR(z)$ of the GRB rate to star formation rate are significant and include
addressing fundamental questions such as incompleteness in rest-frame UV surveys for 
determining the star formation rate at high redshift and time variations in the stellar initial mass function.  
Using an extensive sample of 112 GRBs above a fixed luminosity limit drawn from the 
Second {\Swift} Burst Alert Telescope catalog and accounting for uncertainty in their 
redshift distribution by considering the contribution of ``dark'' GRBs, 
we compare the cumulative redshift distribution $N(<z)$ of GRBs with the star formation
density  $\rhosfr(z)$ measured from UV-selected galaxies over $0<z<$4.  
Strong evolution (e.g., $\GRBSFR(z)\propto(1+z)^{1.5}$ ) is disfavored (Kolmogorov-Smirnov
test $P<0.07$).
We show that more modest evolution (e.g., $\GRBSFR(z)\propto(1+z)^{0.5}$ ) is consistent with the
data ($P\approx0.9$) and
can be readily explained if GRBs occur primarily in low-metallicity
galaxies which are proportionally more numerous at earlier times. If such trends
continue beyond $z\simeq$4, we find the discovery rate of distant GRBs implies a star
formation rate density much higher than that inferred from UV-selected galaxies.  While some previous 
studies of the GRB-star formation rate connection have concluded that GRB-inferred star
formation at high redshift 
would be sufficient to maintain cosmic reionization over 6$<z<$9 and reproduce 
the observed optical depth of Thomson scattering to the cosmic microwave background, we 
show that such a star formation history would over-predict the observed stellar mass density 
at $z>4$ measured from rest-frame optical surveys. The resolution of this important
disagreement is currently unclear, and
the GRB production rate at early times is likely more complex than a simple function of
star formation rate and progenitor metallicity. 
\end{abstract}

\keywords{gamma rays: bursts -- galaxies: evolution}

\section{Introduction}
\label{section:intro}

The history of star formation in the universe is fundamental
for determining the redshift-dependent properties of the galaxy population, 
the production of metals in the intergalactic medium, and the ionization
state of cosmic gas.  Observational probes of star formation
at cosmological distances are therefore valuable tools for
learning about the bulk properties of the universe and
its contents \citep[for a review, see][]{robertson2010a}.

The current frontier in this field concerns the observational determination
of the comoving density of star formation $\rhosfr(z)$ beyond a redshift $z\simeq$6.
This effort is key to understanding whether early galaxies were 
responsible for cosmic reionization as well as when this phase
transition in the intergalactic medium (IGM) occurred. Good progress
has been made through efforts to identify rest-frame 
ultraviolet (UV)-selected galaxies in deep optical and near-infrared imaging
out to $z\sim8$  \citep[e.g.,][]{bouwens2007a,bouwens2010c,bouwens2010a,oesch2010a,mclure2010a}.
Recent spectroscopic campaigns have begun to confirm that many of these
sources do indeed lie at $z\gtrsim7$ \citep[e.g.,][]{iye2006a,stark2010a,lehnert2010a,fontana2010a,vanzella2011a,schenker2011a,pentericci2011a,ono2011a}. 
However, the direct detection of $z>6$ galaxies spanning the full range of luminosities
necessary to reliably deduce the integrated star formation rate represents a challenge
with current facilities. For example, the observationally-inferred $\rhosfr(z)$ depends 
on the adopted value of the poorly-determined luminosity function faint end slope 
and the magnitude limit of the survey \citep{mclure2010a,bouwens2010c}.

As long-duration Gamma Ray Bursts (GRBs) are thought to occur
through the demise of very massive, possibly metal-poor stars 
\citep[for a review, see][]{woosley2006a},
and have been observationally connected to broad-line
Type Ic supernovae \citep[SNe Ic; e.g.,][]{galama1998a,stanek2003a,hjorth2003a},
the rate of high redshift GRBs of this class could provide a valuable,
complementary estimate of  $\rhosfr(z)$ \citep[e.g.,][]{totani1997a,wijers1998a,mao1998a,porciani2001a,bromm2002a,chary2007a}.
In particular, as luminous events, they could effectively probe
the full extent of the star-forming galaxy population including low mass
systems undetected in the deepest optical and near-infrared surveys.
The recent discovery of several $z>6$ long-duration GRBs \citep{kawai2006a,greiner2009a,tanvir2009a,salvaterra2009a,cucchiara2011a}
makes this a particularly interesting avenue to explore.

The connection between the rate of GRBs (comoving Mpc$^{-3}$ yr$^{-1}$)
and the density of star formation ($M_{\odot}$ yr$^{-1}$ Mpc$^{-3}$) can
be coarsely parameterized by  $\GRBSFR(z)$, the ratio of the
GRB rate to the star formation rate density $\rhosfr$. The ratio $\GRBSFR(z)$ 
could be inferred by comparing the cumulative redshift distribution 
$N(<z)$ of GRBs with the observed evolution of the star-formation rate
density $\rhosfr(z)$ over the range of redshifts where $\rhosfr(z)$
is well-measured. Earlier work  \citep{yuksel2008a,kistler2008a,kistler2009a,wyithe2010a}
has suggested that this ratio increases with redshift in the sense that
GRBs are more frequent for a given star formation rate density at earlier times.
For instance, if such a redshift dependence is parameterized as a simple power law, 
$\GRBSFR(z)\propto(1+z)^{\alpha}$, \citet{kistler2009a} find 
that $\alpha\sim1.2$.  Although the origin of such a redshift-dependence remains unclear, 
if extrapolated to higher redshifts the abundance of GRBs at $z>7$ 
\citep{tanvir2009a,salvaterra2009a,cucchiara2011a} implies a density
of star formation considerably higher ($\sim6-7\times$ larger) than 
$\rhosfr(z)$ measured from the abundance of distant UV-selected galaxies.
In other words, the decline in the GRB rate at $z>4$ is considerably less rapid than
the decrease in the star formation rate \citep[e.g.,][]{wanderman2010a}.
This discrepancy has potentially important implications for the reionization 
history of the universe and particularly the contribution from early star-forming
galaxies \citep[e.g.,][]{kistler2009a,wyithe2010a}.

Previous determinations of $\GRBSFR(z)$ at low to intermediate
redshift \citep[e.g.,][]{yuksel2008a,kistler2008a,kistler2009a,ishida2011a}
have not accounted for the uncertainty in the GRB redshift distribution $N(<z)$ 
that arises from the so-called ``dark'' GRBs defined as those with absent or faint optical afterglows.
\citet{perley2009a} have shown that dark GRBs likely span a wide range of 
redshifts and so they can be used to provide a valuable estimate of the uncertainty in the overall
distribution, corresponding to approximately $25\%$ in $N(<z)$
at fixed $z$; see their Figure 8 (see also, e.g., \citealt{greiner2011a} and \citealt{kruhler2011a}).  
In this work we use the recently-released 
Second {\Swift} Burst Alert Telescope (BAT) Catalog of GRBs \citep{sakamoto2011a}
to compile a comprehensive sample of 112 luminous ($\Liso > 10^{51}$ ergs s$^{-1}$) 
long-duration GRBs with known redshifts and upper limits
\citep{butler2007a,butler2010a,fynbo2009a,sakamoto2011a}, accounting for
the uncertainty in the overall distribution by including constraints provided 
for a representative subsample of dark GRBs \citep{perley2009a,greiner2011a,kruhler2011a}.

Our goal is to revisit the redshift-dependent GRB rate to star-formation
rate ratio $\GRBSFR(z)$ with this improved dataset to determine
carefully whether the ratio evolves and if so on what physical basis. A likely
driver of such evolution is metallicity in the host galaxy population. 
To explore this possibility, we examine whether the
evolution of the stellar mass-metallicity \citep[e.g.,][]{tremonti2004a,savaglio2005a,erb2006a}
and star formation-stellar mass \citep[e.g.,][]{drory2008a}
relations, when combined with the redshift-dependent stellar mass function
\citep[e.g.,][]{drory2005a}, can explain the redshift scaling of $\GRBSFR(z)$.

Our physical analysis of $\GRBSFR(z)$ over 0$<z<4$ enables us to 
calculate the high-redshift ($4<z<10$) star formation rate density implied 
by the presence of distant GRBs and to compare this $\rhosfr(z)$ with estimates drawn
from rest-frame UV galaxy surveys. Following arguments developed in \citet{robertson2010a}
we calculate the impact of this GRB-deduced star formation
rate density on other indicators of the reionization history, including
the optical depth of electron scattering to the cosmic microwave background
and the stellar mass density that represents the integral of earlier activity. 
We show these additional constraints provide a critically important boundary 
on what is physically plausible in terms of the
GRB-derived star formation rate at $z>6$ in addition to the likelihood of
the various models we develop to explain the redshift dependence of
 $\GRBSFR(z)$.  We conclude by summarizing the issues 
that will need reckoning before the connection between the high-redshift 
GRB and star formation rates is understood.

This paper is organized as follows. We construct a robust sample of GRBs
above a fixed luminosity limit and discuss their redshift distribution and its
uncertainties in \S \ref{section:observations}.  In \S \ref{section:comparison}, we
compare the observed cumulative redshift distribution of GRBs over $0<z<$4 to the
equivalent star formation rate density and its evolution. This comparison enables us
to consider whether and how the GRB rate to star formation rate
ratio $\GRBSFR$ evolves. In \S \ref{section:model} we interpret our results
in the context of a hypothesis where GRBs primarily occur
in low-metallicity galaxies. We compare predictions of this model with the
observed GRB redshift distribution.
In \S \ref{section:implications}, we then use the observed GRB rate beyond z$\simeq$4 
to infer the star formation rate density under various assumed forms for $\GRBSFR(z)$.
We discuss the ramifications of our results in the context of other constraints on cosmic reionization
in \S \ref{section:reionization}.

Throughout the paper we have assumed a standard flat $\Lambda$CDM cosmology 
($\Omega_{m}=0.3$, $\Omega_{\Lambda}=0.7$, $h=0.7$).

\section{The Redshift Distribution of Long Duration GRBs}
\label{section:observations}

Our goal is to construct a robust sample of GRBs, observed down to a fixed luminosity
limit, within a redshift range where a meaningful comparison can be
made with an independently-determined star formation rate density, $\rhosfr(z)$,
from rest-frame UV surveys of galaxies. In this way, the connection between the GRB rate and the
star formation rate density, $\GRBSFR$, and its possible evolution can be studied.

In a manner similar to \citet{kistler2009a}, we can describe
the observable number of GRBs within a redshift range 
$\za\leq z\leq\zb$ as
\begin{equation}
\label{eqn:N_GRB}
N(\za,\zb) = K \int_{\za}^{\zb} \rhosfr(z) \GRBSFR(z) \frac{\mathrm{d}V}{\mathrm{d}z}\frac{\mathrm{d}z}{1+z} 
\end{equation}
\noindent
where $\rhosfr(z)$ is the global star formation rate density,
$\GRBSFR$ is the number of GRBs per unit star formation rate,
$\mathrm{d}V/\mathrm{d}z$ is the redshift-dependent volume
element, and the factor $1/(1+z)$ accounts for cosmological
time dilation affecting the observed GRB rate. The constant
$K$ is an factor that accounts for the efficiency of the GRB 
search (e.g., areal coverage, the survey flux limit, etc.), but
its value is unimportant for our purposes.\footnote{A clear concern is the
possible multiplicative degeneracy between $K$ and $\GRBSFR(z)$ in Equation
\ref{eqn:N_GRB}, and that possible redshift dependence in, e.g., the follow-up
efficiency could mimic an evolution of $\GRBSFR(z)$. We note that \cite{dai2009a}
have found that the cumulative peak photon flux distribution of {\it Swift}
GRBs with and without spectroscopic redshifts are similar.  This result 
suggests that the follow-up efficiency does not strongly depend on redshift.}
 We can remove the
dependence on $K$ by simply constructing the cumulative redshift
of GRBs over the redshift range $0<z<\zlim$, normalized to $N(0,\zlim)$,
as
\begin{equation}
\label{eqn:Nz}
N(<z|\zlim) = \frac{N\left(0,z\right)}{N\left(0,\zlim\right)}
\end{equation}
\noindent
The product $\rhosfr(z)\GRBSFR(z)$ then sensibly determines the cumulative
redshift distribution of GRBs.  Given this normalization, we will also
sometimes refer to the redshift-dependent fraction of star formation that
can produce GRBs $\GRBSFRl(z)$, which is related to the number of GRBs per
unit star formation rate as $\GRBSFR(z) = \GRBSFR_{0} \GRBSFRl(z)$.  The
constant $\GRBSFR_{0}$ then encodes the number of GRBs formed per unit mass of
stars.  The value of $\GRBSFR_0$ cannot be determined independently of the unknown
$K$ in Equation \ref{eqn:N_GRB}, but the time or redshift when $\GRBSFR_0$ is 
defined does matter
for a model where $\GRBSFRl(z)$ is calculated directly (see \S \ref{section:model} 
below) rather than averaged over some redshift interval (c.f., Equation \ref{eqn:Nz}). In
such a case, we take $\GRBSFR_0$ to be defined relative to when $\GRBSFRl(z) = 1$.

\subsection{GRB Catalog}
\label{section:observed_grbs}

To evaluate the possibility of redshift-dependence in the 
GRB to star formation rate density $\GRBSFR(z)$ through Equation
\ref{eqn:N_GRB}, we require an observational sample to construct the cumulative redshift distribution
calculated by Equation \ref{eqn:Nz}. The primary requirements are a well-understood
completeness in the redshift determinations above some GRB isotropic-equivalent luminosity.  
The distribution
$N(<z|\zlim)$ can be determined from catalogs of GRBs monitored with gamma-ray satellites 
\citep[e.g., {\it Swift};][]{gehrels2004a} and followed up 
from the ground \citep[e.g.,][]{fynbo2009a}.  For our base catalog of GRBs,
we take the union of the samples presented in \citet{butler2007a}, \citet{perley2009a},
\citet{butler2010a}, \citet{sakamoto2011a}, \citet{greiner2011a}, and \citet{kruhler2011a}. 
We include only GRBs occurring before the end of the Second {\it Swift} BAT GRB Catalog and
prefer the most recent redshifts for GRBs where
the samples disagree.  This union provides a sample of 164 GRBs with known redshifts and redshift
upper limits, but two
GRBs (071112C and 060505) have incomplete fluence or burst duration measures and are discarded. 
The remaining 162 long duration GRBs with redshifts or redshift limits serve as our base GRB catalog.

To account for the incompleteness
owing to the flux limit of the {\it Swift} survey, we follow \citet{kistler2009a} 
and construct a subsample of with isotropic-equivalent luminosities $\Liso>10^{51}$ ergs s$^{-1}$. 
The luminosity is computed as
\begin{equation}
\label{eqn:L_iso}
\Liso \equiv \frac{\Eiso}{t_{90}/(1+z)}
\end{equation}
\noindent
where $\Eiso$ is the isotropic-equivalent energy, 
$t_{90}$ is the burst duration containing from 5\% to 95\% of the total fluence,
 and the factor of $(1+z)$ accounts for cosmological time dilation \citep[e.g.,][]{kistler2009a}.  
For isotropic-equivalent energies, 129 GRB $\Eiso$ values are taken from
\citet{butler2007a,butler2010a} and 4 recent values (060908, 090926B, 091018, and 091029)
from \citet{sakamoto2011a}.  
Additionally, the isotropic energy values for
21 further GRBs are computed from the fluences reported by \citet{butler2007a,butler2010a}.
The isotropic energies from the 8 remaining GRBs 
(060512, 090814A, 090904B, 090927, 091020, 091024, 091127, and 091208B)
are calculated from the 15-150 keV
fluences reported by 
\citet[][]{sakamoto2011a}, 
but are possibly lower limits given the 
typical energy range of $1-10000$keV for defining isotropic-equivalent quantities 
\citep[e.g.,][]{amati2002a}.  The burst durations $t_{90}$ are taken from Table 1 of
\citet[][]{sakamoto2011a}, except for GRBs 050820A, 060218, and 090529A taken from 
\citep{butler2007a,butler2010a}.
For our catalog of 162 GRBs with redshifts and redshift limits, this culling
provides 112 GRBs with isotropic-equivalent 
luminosities $\Liso>10^{51}$ ergs s$^{-1}$ for our analysis.  The redshifts and limits, isotropic 
equivalent
energies and luminosities, and burst durations of the full sample of 162 GRBs
compiled from the union of the \citet{butler2007a}, \citet{perley2009a}, \citet{butler2010a}, 
\citet{sakamoto2011a}, \citet{greiner2011a} and \citet{kruhler2011a} catalogs are provided for convenience
in Table \ref{table:grbs} in the Appendix.

Since we will use the cumulative redshift distribution $N(<z)$ of this sample as the basis for our analysis, it
is important to consider its uncertainties. While the {\it Swift} catalogs provide a valuable compilation of 
gamma-ray detections, the redshift determinations are clearly influenced by their optical 
observability.  
The phenomenon of so-called "dark" GRBs with suppressed optical counterparts could influence whether 
the observed $N(<z)$ is representative of that for {\it all} long-duration GRBs.  
\citet{perley2009a} have considered
this important issue by attempting to constrain the redshift distribution 
of dark GRBs through deep
searches that successfully located faint optical and near-infrared counterparts.  
The \citet{perley2009a} work provides us with 2 redshifts 
and 9 redshift upper limits for a subsample of dark GRBs in our catalog.
\citet{greiner2011a} and
\citet{kruhler2011a} have pursued this effort in parallel, and have provided 3 additional redshifts and
1 redshift upper limit for dark GRBs in our catalog.  We assume 
the subsamples of dark GRBs with redshift upper limits presented by  
\citet{perley2009a}, \citet{greiner2011a}, and \citet{kruhler2011a} 
are representative of that class, and therefore optionally incorporate
those limits to characterize the effects of possible 
incompleteness of the {\it Swift} sample with firm redshift determinations.
We also note that while the luminosity limit for our sample was chosen to 
match \citet{kistler2009a}, at redshifts above $z>4$ the {\it Swift} sample
is incomplete for this limit.  However, fully accounting for this 
incompleteness would only increase the relative number of GRBs at high-redshifts.
As our following results show, our sample's luminosity limit is therefore conservative.

\section{Comparing GRB Rates to the Cosmic Star Formation History}
\label{section:comparison}

As we have yet to develop 
physical intuition into the connection between the rate of GRBs and $\rhosfr(z)$, we will begin
by an empirical comparison of the cumulative GRB redshift distribution $N(<z|\zlim)$ constructed as described 
in \S \ref{section:observed_grbs} with the cumulative redshift distribution that would be expected given the 
observed star formation rate density $\rhosfr(z)$ from rest-frame UV-selected sample and various forms for
the redshift-dependent ratio $\GRBSFR(z)$.

For $\rhosfr(z)$ we use the results from \citet{hopkins2006a} who gathered and standardized
measures of the star formation rate density from the \citet{hopkins2004a} compilation and observations by \citet{wolf2003a},
\citet{bouwens2003a,bouwens2003b}, \citet{bunker2004a}, \citet{ouchi2004a}, \citet{arnouts2005a}, \citet{lefloch2005a},
\citet{perez-gonzalez2005a}, \citet{schiminovich2005a}, \citet{bouwens2006a}, \citet{hanish2006a}, 
and \citet{thompson2006a}.  \citet{hopkins2006a} provide a piecewise-linear ``Modified Salpeter A IMF'' model in their 
Table 2 that provides a good statistical fit to  the available star formation density data.  We limit our use of their fit to $z<4$ 
where the data is optimal.  We note that $\rhosfr(z)$ is relatively flat  
[$\rhosfr(z)\propto(1+z)^{-0.26}$] in the interval $1<z<4$ 
where most of the GRBs with $\Liso>10^{51}$ ergs s$^{-1}$ occur. This scaling
means the use of Equation \ref{eqn:Nz} is particularly
accurate as the factor of $\sim2$ in the normalization of $\rhosfr(z)$ allowed by the $3-\sigma$ 
uncertainty in the \citet{hopkins2006a} 
fit is circumvented (see also \S \ref{section:model} below).  We note that using the 
\citet{hopkins2006a} results for the
\citet{baldry2003a} IMF instead of the Salpeter A IMF model \citep[see, e.g.,][]{hopkins2008a} 
has little effect on our conclusions.

Figure \ref{fig:Nz} shows the cumulative redshift distribution of observed GRBs (black histogram), normalized over the
redshift range $0<z<4$.  The gray-shaded region shows how the distribution shifts in the limiting cases of all dark GRBs 
occurring at $z=0$ or the upper redshift limits determined by \citet{perley2009a}, \citet{greiner2011a}, and \citet{kruhler2011a}.
Figure \ref{fig:Nz} compares the observed GRB cumulative redshift distribution $N(<z|\zlim=4)$ for the case of three
models for the redshift evolution of the GRB rate to SFR ratio $\GRBSFR(z)$.  

If the quantity $\GRBSFR(z)\sim$ is constant (red line), the cumulative redshift distribution of GRBs increases rapidly at $z\sim2-3$ in sync with
the star formation rate density.  If instead the GRB rate to SFR ratio evolves strongly with redshift over the
epoch $z<4$ as, for instance, $\GRBSFR(z)\propto(1+z)^{1.5}$ (orange line), then the GRB rate is shifted to higher
redshifts and the cumulative distribution increases rapidly at $z>3$.  A weaker redshift evolution [$\GRBSFR(z)\propto(1+z)^{0.5}$] 
better reproduces the cumulative GRB rate density (blue line). Given the relatively small sample, the data appears 
roughly consistent with each of these $\GRBSFR(z)$ redshift-scalings.

\subsection{Statistical Tests}
\label{section:tests}

Given the integral distribution of observed GRB redshifts, and a parameterized model for predicting this distribution,
we can perform statistical tests to check for consistency between the observed and model distributions, and calculate
confidence intervals for the parameters of the model given the data.  The null hypothesis that the observed GRB
redshifts are consistent with a model distribution can be evaluated by the one-sample Kolmogorov-Smirnov (KS) test, 
which calculates a "$P$-value" that corresponds to one minus the probability that the null 
hypothesis can be rejected.
We employ this statistical test to determine a plausible range of values for the paramter $\alpha$ in a model where
$\GRBSFR(z)\propto(1+z)^{\alpha}$.  We only
use the KS test to conservatively evaluate the relative agreement between the
observed and model GRB redshift distributions.  More formally, we can calculate the likelihood function for the
parameter $\alpha$ given the observed data and assumed parameterized model using a Bayesian technique.  For the
case of $N$ independent samples $\mathbf{z} = [z_{i}]$ from a redshift probability distribution $p(\mathbf{z}|\alpha)$ (e.g., the 
integrand of Equation \ref{eqn:N_GRB}), the likelihood function
of $\alpha$ can be calculated as
\begin{equation}
\label{eqn:likelihood}
\mathcal{L}(\alpha) = \prod_{i=0}^{N-1} p(z_{i}|\alpha).
\end{equation}
\noindent
The posterior probability density of $\alpha$ given the observed data $z_{i}$ and parameterized model $\GRBSFR$ can then be computed as
\begin{equation}
\label{eqn:posterior}
p(\alpha|\mathbf{z},\GRBSFR) = \frac{p(\alpha|\GRBSFR) \mathcal{L}(\alpha)}{\int \mathrm{d}\alpha p(\alpha|\GRBSFR) \mathcal{L}(\alpha)}
\end{equation}
\noindent
where $p(\alpha|\GRBSFR)$ is the prior probability of the parameter $\alpha$ given the model $\Psi$.  We take this prior to be flat
over a wide range of $\alpha$ such that it does not affect the shape of $p(\alpha|\mathbf{z},\GRBSFR)$.  The confidence interval
corresponding to a given Gaussian equivalent significance can then be calculated by integrating Equation \ref{eqn:posterior} about the
peak likelihood, and whenever an effective $\sigma$ is quoted it refers to the Bayesian confidence region.

\begin{figure}
\figurenum{1}
\epsscale{1}
\plotone{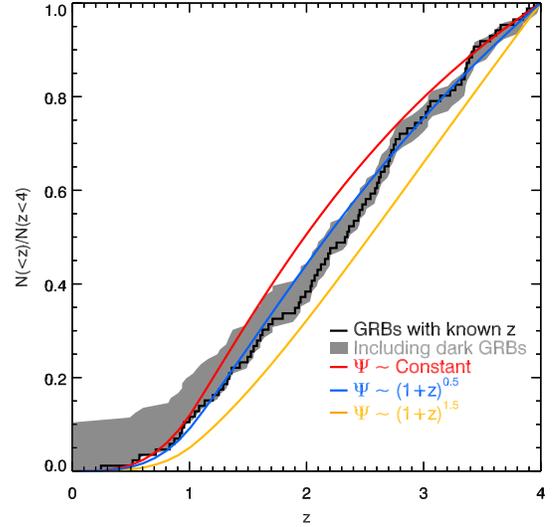}
\caption{\label{fig:Nz}
\scriptsize
Cumulative redshift distribution of long-duration gamma-ray bursts (GRBs) at $z<4$ determined from the second {\it Swift} 
BAT catalog with isotropic equivalent luminosities $\Liso>10^{51}$ ergs s$^{-1}$ (black histogram, see text for source of
data).  Incorporating the redshift constraints of dark GRBs shifts the distribution over the range indicated 
by the gray shaded area, and provides a estimate of the intrinsic uncertainty in the observational determination of the distribution.  
Three simple power-law parameterizations of the ratio $\GRBSFR(z)$ between the GRB and star formation rate densities are shown: a constant ratio with
redshift (red line), $\GRBSFR(z)\propto(1+z)^{0.5}$ (blue line), and $\GRBSFR(z)\propto(1+z)^{1.5}$.
}
\end{figure}

\begin{figure}
\figurenum{2}
\epsscale{1}
\plotone{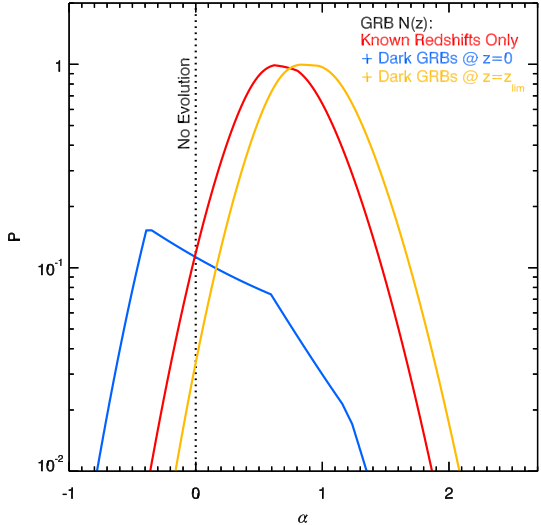}
\caption{\label{fig:KS}
\scriptsize
Results of a Kolmogorov-Smirnov test comparing the GRB cumulative redshift distribution 
$N(<z|\zlim=4)$
with model distributions calculated 
assuming the \citet{hopkins2006a} model for the star formation rate density $\rhosfr(z)$ with a 
redshift-dependent GRB rate to SFR ratio $\GRBSFR(z)\propto(1+z)^{\alpha}$.  Shown is the 
probability $P$ of consistency as a function of the power-law index $\alpha$.
Comparing with sample with spectroscopic redshifts (red line), the $P>0.05$ interval is 
$-0.2<\alpha<1.5$ 
with a peak probability near $\alpha=0.5$.  The Gaussian equivalent $2-\sigma$ confidence interval 
is $0.025<\alpha<1.5$ (see text).
If the dark GRB sample with redshift upper limits is assumed to be local ($z\approx0$), the
$P>0.05$ interval shifts to $-0.7<\alpha<1$ with a median near $\alpha\approx0$. If the dark GRBs 
lie at their maximum possible redshift the interval moves to $0<\alpha<1.8$ with a peak near 
$\alpha\approx0.8$.   
}
\end{figure}

Figure \ref{fig:KS} shows the KS-test probability assuming 
$\GRBSFR(z) \propto (1+z)^{\alpha}$ for  $-1<\alpha<2.5$.  Comparing only with GRBs with spectroscopic redshifts (the solid black
histogram in Figure \ref{fig:Nz}), we find that the region where KS $P>0.05$
contains power-law indices
of $-0.2 \lesssim\alpha\lesssim 1.5$.  The peak probability occurs for $\alpha\approx0.5$.  Computing the posterior probability $p(\alpha|\mathbf{z},\GRBSFR)$,
we find a constant
$\GRBSFR$ (i.e., no evolution)
 is marginally allowed at the $2-\sigma$ level.  Including the dark GRB redshift constraints 
shifts the probability curve.  If all dark GRBs are local, then the KS test region $P>0.05$ contains power-law indices 
$-0.7<\alpha<1.0$ (with a median probability near $\alpha=0$ corresponding to no evolution).  Instead, 
if all dark GRBs are at their maximum possible redshift then within $2-\sigma$ 
power-law indices $0.3<\alpha<1.7$ are favored.  While the constraining power of 
the current sample is not particularly stringent, further monitoring of GRBs should soon definitively rule-out a $\GRBSFR(z)$ 
that is constant or declines with redshift.  We discuss the implications of this constraint further in \S 
\ref{section:implications}.

In summary, we find that the distribution of GRBs with spectroscopic redshifts is consistent with 
only moderate variation of $\GRBSFR(z)$ over $0\lesssim z \lesssim 4$ and that there is overall only
weak evidence for evolution ($\sim2\sigma$ confidence). Compared to previous studies 
\citep[e.g.,][]{kistler2009a}
the results are consistent at the $\sim2\sigma$-level, but we infer a weaker redshift dependence owing
to the fractionally increased number of GRBs at $z\lesssim2$ in our compiled GRB sample.
The additional uncertainty arising from including constraints from dark GRBs is important to include.
If dark GRBs occur at their maximum allowed redshifts the distribution is more heavily weighted towards 
higher redshifts, and more strongly indicates a possible redshift dependence to $\GRBSFR(z)$.  

Clearly, to make progress it would be helpful to have a physical basis for evolution in $\GRBSFR(z)$. 
The most likely hypothesis links the GRB production rate with the metallicity of the underlying stellar
population \citep[e.g.,][]{woosley2006a}. Since the typical metallicity of galaxies at fixed stellar mass 
declines with redshift \citep[e.g.,][]{tremonti2004a,savaglio2005a,erb2006a} this hypothesis provides a natural
basis for an increase in $\GRBSFR(z)$.

\section{Modeling the Redshift-Dependent GRB to SFR Ratio with Metallicity Evolution}
\label{section:model}

We now consider a physical scenario where the cosmic GRB rate is enhanced in low metallicity galaxies, and 
develop a method for calculating the fraction of the star formation rate density occurring below
a characteristic metallicity $\Zcrit$ above which GRB production is suppressed.  Using 
the same statistical comparisons employed in \S \ref{section:comparison}, we can evaluate
whether the observed GRB rate is consistent with such a picture and, if so, what values
of $\Zcrit$ are favored.

A variety of theoretical pictures for the origin of long duration GRBs \citep[e.g.,][]{woosley1993a} suggest that the
GRB rate may be connected to the metallicity of their progenitor stellar population \citep[for reviews, see]
[]{meszaros2006a,woosley2006a}.  GRBs may require the retention of significant angular momentum 
after collapse, which limits the amount of mass loss prior to collapse \citep[e.g.,][]{fryer1999a,macfadyen2001a}.  
The lower opacity and mass loss rates of low-metallicity stars eases this requirement and provides a credible 
rationale for preferring low-metallicity GRB hosts \citep{hirschi2005a,yoon2005a,woosley2006b}.
Moreover, host galaxies of long duration GRBs are often observed to be metal poor and have low luminosities, 
both in the nearby universe \citep[e.g.,][]{prochaska2004a,sollerman2005a,modjaz2006a,stanek2006a,wiersema2007a}
and at cosmological distances \citep[e.g.,][]{fruchter1999a, fruchter2006a, lefloch2003a, fynbo2003a, savaglio2009a}.  

The quantitative details of the metallicity connection are still debated 
\citep[for a good discussion, see][]{modjaz2011a}.  For instance, by exploiting the connection 
between SNe Ic displaying broad lines and GRBs \citep[e.g.,][]{galama1998a,stanek2003a,hjorth2003a}, \citet{modjaz2008a} 
found that only low-metallicity ($12 + \log[\mathrm{O}/\mathrm{H}] < 8.7$ on the \citealt{kobulnicky2004a} scale,
converted following \citealt{kewley2008a})
galaxies have GRBs that fade into SNe Ic, whereas host galaxies with metallicities above this ceiling 
have GRB-free SNe Ic.  However, there are regions within GRB host galaxies known with higher metallicities 
\citep{levesque2010a}, and the radio-relativistic SN 2009bb \citep{soderberg2010a} occurred in a
high-metallicity region of its host galaxy \citep{levesque2010c}. 
\citet{han2010a} and \citet{levesque2010b} find that GRB host galaxies lie 
below the \citet{tremonti2004a} stellar mass-metallicity relation, whereas \citet{savaglio2009a} report
that the properties of GRB hosts do not clearly differ from normal star-forming galaxies.  
\citet{mannucci2011a} find that GRB hosts are offset from the mass-metallicity relation owing to their
their higher star formation rates than average at fixed stellar mass, but remain within the ``fundamental
metallicity relation'' between star formation rate, stellar mass, and metallicity \citep{mannucci2010a}.

Given this controversy, we considered it appropriate to construct a model to calculate $\GRBSFR(z)$ in the case 
where GRBs preferentially occur in host galaxies with metallicities below some characteristic $\Zcrit$.
\citet{kocevski2009a} combined the \citet{modjaz2008a} empirical host galaxy metallicity ceiling for GRBs with the
redshift evolution of the stellar mass-metallicity relation \citep{savaglio2005a}, the stellar mass-
star formation rate relation evolution \citep{drory2008a}, and the stellar mass function evolution 
\citep{drory2005a} to calculate the redshift-dependent characteristic mass of GRB host galaxies.
Below, we extend the \citet{kocevski2009a} formalism to model the redshift dependence of $\GRBSFR(z)$ and
allow for a variable metallicity ceiling $\Zcrit$. We note that a similar approach has been adopted by \citet{langer2006a},
\citet{salvaterra2007a}, \citet{salvaterra2009b}, \citet{butler2010a} and \citet{virgili2011a}. Analyses that used the
\citet{langer2006a} calculation utilized the fraction of stellar mass at metallicities below $\Zcrit$ to determine the redshift-dependence of 
$\GRBSFR(z)$, whereas we prefer to extend the \citet{kocevski2009a} model to find $\GRBSFR(z)$ from the fraction of
{\it star formation} occurring at metallicities below $\Zcrit$.

\subsection{A Model for $\GRBSFR(z)$ from Metallicity Evolution}
\label{section:kocevski}

Given the relation between stellar mass and metallicity \citep[e.g.,][]{tremonti2004a} a
metallicity ceiling $\Zcrit$ for GRB host galaxies would imply a critical galaxy stellar mass $\Mcrit$ 
above which GRB production is suppressed.  The redshift-dependent stellar-mass metallicity relation 
can be parameterized as \citep[][on the \citealt{kobulnicky2004a} scale]{savaglio2005a}
\begin{eqnarray}
\label{eqn:Z_vs_m}
12 + \log[\mathrm{O}/\mathrm{H}] &=& -7.5903 + 2.5315\log \Mstar\nonumber\\
&& -0.09649 \log^{2} \Mstar\nonumber\\
&& +5.1733\log t_{u} - 0.3944\log^{2} t_{u}\nonumber\\
&& -0.403\log t_{u} \log \Mstar,
\end{eqnarray}
\noindent
where $t_{u}$ is the age of the universe at redshift $z$ in Gyr and $\Mstar$ is the galaxy stellar mass in 
solar masses.  Equation \ref{eqn:Z_vs_m} then can be used to connect a given metallicity ceiling $\Zcrit$ to a
critical galaxy mass $\Mcrit(z)$ for GRB production.

We now introduce the fraction of star formation occurring in galaxies with metallicities lower than $\Zcrit$, which can be
expressed as
\begin{equation}
\label{eqn:GRBSFR}
\GRBSFRl(z) = \frac{ \int_{0}^{\Mcrit(z)} \SFR(M,z) \Phistar(M,z) \mathrm{d} M}{ \int_{0}^{\infty} \SFR(M,z) \Phistar(M,z) \mathrm{d} M}
\end{equation}
where $\SFR(M,z)$ is the star formation rate - stellar mass relation and $\Phistar(M,z)$ is the galaxy stellar mass
function.  \cite{drory2008a} parameterize the observed star formation rate-stellar mass relation as
\begin{equation}
\label{eqn:sfr_vs_m}
\SFR(M,z) = \SFR_{0}\left(\frac{M}{M_{0}}\right)^{\beta} \exp\left(-\frac{M}{M_{0}}\right)
\end{equation}
\noindent
where $\beta\approx0.5$ and the parameters $\SFR_{0}$ and $M_{0}$ evolve with redshift as
\begin{eqnarray}
\label{eqn:sfr_vs_m_params}
\SFR_{0}(z) = 3.01 (1+z)^{3.03} M_{\sun}\,\,\mathrm{yr}^{-1}\\
M_{0}(z)    = 2.7\times10^{10}(1+z)^{2.1} M_{\sun}.
\end{eqnarray}
\noindent

The stellar mass function also evolves with redshift \citep[e.g.,][]{drory2005a}.  Taking
a \cite{schechter1976a} model for the galaxy stellar mass
\begin{equation}
\label{eqn:phistar}
\Phistar(M,z) = \phistar\left(\frac{M}{M_{1}}\right)^{\gamma}\exp\left(-\frac{M}{M_{1}}\right) \frac{\mathrm{d}M}{M_{1}},
\end{equation}
\noindent
\cite{drory2008a} model the redshift-dependence of the observed galaxy stellar mass function through the
parameters
\begin{equation}
\phistar(z) \approx 0.003(1+z)^{-1.07}\,\,\mathrm{Mpc}^{-3} \mathrm{dex}^{-1}\\
\end{equation}
\begin{equation}
\log[M_{1}/M_{\sun}](z) \approx 11.35 - 0.22\ln(1+z)\\
\end{equation}
\begin{equation}
\label{eqn:phistar_params}
\gamma(z) \approx -1.3.
\end{equation}

By combining Equations \ref{eqn:Z_vs_m} and \ref{eqn:sfr_vs_m}-\ref{eqn:phistar_params} the
redshift-dependent ratio of the GRB rate to the star formation rate $\GRBSFR(z)$ can be estimated by evaluating
Equation 5.  With $\GRBSFR(z)$ in hand, a model for the cumulative redshift distribution of GRBs can be 
calculated using Equations \ref{eqn:N_GRB} and \ref{eqn:Nz} once a model of the star formation rate
density $\rhosfr(z)$ is adopted.  For a critical host galaxy
metallicity $\Zcrit$ above which GRBs are suppressed, we can use the GRB data to inform us as to what
metallicity ceilings are plausible.  The redshift-dependence of $\GRBSFR(z)$ will clearly depend on
the value of $\Zcrit$ since the fraction of star formation occurring at metallicities below $\Zcrit$
will vary with redshift owing to the evolution of the mass-metallicity relation, the mass-star formation rate
relation, and the stellar mass function. For simplicity in this model, GRBs are prevented from occuring 
above $\Zcrit$, but as noted above in \S \ref{section:model} GRBs do occur in metal rich host galaxies.  The
suppression in metal rich galaxies should therefore be understood to be an incomplete, coarse model to 
indicate the potential preference for GRBs to occur in metal-poor systems.

\begin{figure}
\figurenum{3}
\epsscale{1}
\plotone{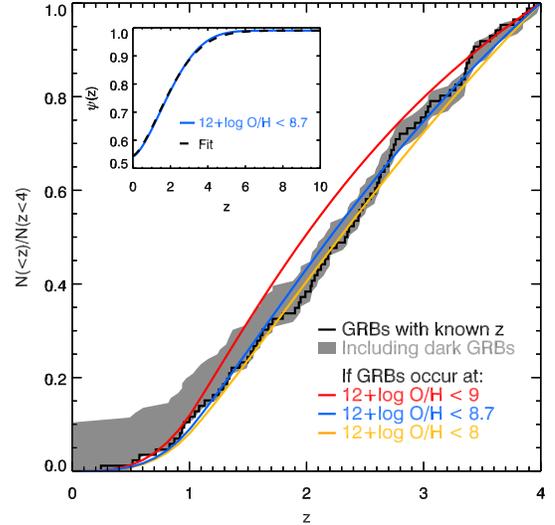}
\caption{\label{fig:Nz.Z}
\scriptsize
Cumulative redshift distribution of gamma-ray bursts at $z<4$ and metallicity-evolution models for 
the GRB rate to star formation rate ratio $\GRBSFR(z)$.  The black histogram and gray area indicate 
the cumulative distribution of GRBs with firm redshifts and the uncertainty owing to dark GRBs (see 
Figure \ref{fig:Nz} caption). Also shown are three models of the ratio $\GRBSFR(z)$ where GRB 
production is suppressed in galaxies with metallicities above a ceiling $\Zcrit$, determined by 
model presented in \S \ref{section:model}: a high $\Zcrit=9$ (red line), the $\Zcrit\approx8.7$ 
ceiling found by \citet[][blue line]{modjaz2008a}, and a illustrative low ceiling $\Zcrit=8$ 
(orange line).  The inset shows the redshift-dependence $\GRBSFRl(z)$ corresponding to the model 
with $\Zcrit\approx8.7$, along with a parameterized fit (see Equation \ref{eqn:GRBSFR_fit}).
}
\end{figure}

Figure \ref{fig:Nz.Z} shows the cumulative GRB redshift distribution $N(<z|\zlim=4)$ resulting from Equation $\ref{eqn:GRBSFR}$ with the adoption of the \cite{hopkins2006a} star formation rate density model, for three choices for the critical metallicity $\Zcrit$.  One model adopts a large $\Zcrit=9$
(red line), similar to the metallicity of the host of GRB 020819 \citep{levesque2010a}.  In this case, essentially all star formation occurs in hosts with metallicities below
$\Zcrit$ and $\GRBSFR$ is roughly constant with redshift as the GRB rate and star formation rate density trace one another.  The large $\Zcrit$ model therefore closely tracks the $\alpha=0$ model shown in Figure \ref{fig:Nz}.  An intermediate model adopts the value of $\Zcrit\approx8.7$
from \citet{modjaz2008a}, shown as the blue line in the Figure \ref{fig:Nz.Z}.  Star formation occurring in galaxies
with metallicities below the \citet{modjaz2008a} $\Zcrit$ tracks the GRB rate with surprising fidelity, and   
for convenience we show the corresponding $\GRBSFR(z)$ in the Figure \ref{fig:Nz.Z} inset and provide a 
parameterized fit to this $\GRBSFR(z)$ model as
\begin{equation}
\label{eqn:GRBSFR_fit}
\GRBSFR_{\mathrm{fit}}(z) = 0.5454 + (1 - 0.5454)\times\left[\mathrm{erf}\left(0.324675z\right)\right]^{1.45}
\end{equation}
\noindent
(dashed black line in Figure \ref{fig:Nz.Z} inset) that recovers the computed $\GRBSFRl(z)$ to within 1\% at $0<z<10$.
Normalized over the redshift range $0<z<4$, this intermediate $\Zcrit$ model produces a cumulative redshift distribution similar 
to the $\GRBSFR(z) \propto (1+z)^{0.5}$ model discussed in \S \ref{section:comparison}.  Third, we show the effects of a low
value of $\Zcrit=8$.  While the fraction of star formation occurring in systems with $12+\log[\mathrm{O}/\mathrm{H}]<8$ is much smaller than
$12+\log[\mathrm{O}/\mathrm{H}]\approx8.7$, normalized over the redshift range $0<z<4$ the redshift dependence of the two models are similar.  
Sensibly, the low $\Zcrit$ model evolves with a somewhat stronger redshift dependence as the epoch at where the
characteristic stellar mass in the stellar mass function reaches $\Zcrit$ is pushed to higher redshift.  This
low $\Zcrit$ model only serves as a strawman to illustrate the calculated redshift dependence; 
GRBs are observed to occur at larger metallicities \citep[see, e.g.,][]{levesque2010b,mannucci2011a}.

\begin{figure}
\figurenum{4}
\epsscale{1}
\plotone{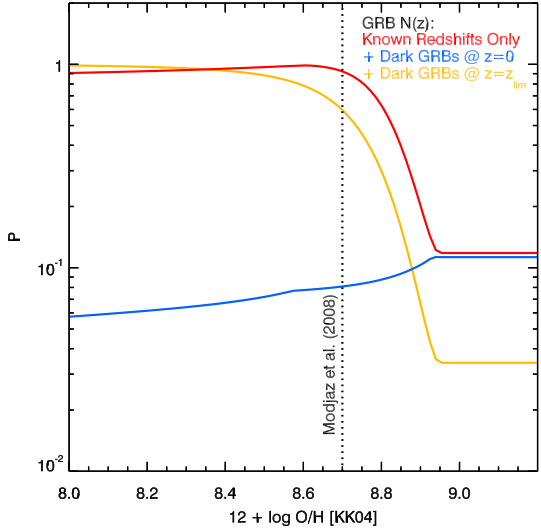}
\caption{\label{fig:KSZ}
\scriptsize
Kolmogorov-Smirnov test results which compare the observed GRB cumulative redshift distribution 
with the calculated distribution $N(<z|\zlim=4)$ assuming the \citet{hopkins2006a} model for the 
star formation rate density $\rhosfr(z)$ with a redshift-dependent GRB rate to SFR ratio 
$\GRBSFR(z)$ given by Equation \ref{eqn:GRBSFR}.  Shown is the probability $P$ of consistency as a 
function of the critical metallicity $\Zcrit$ \citep[on the scale of ][]{kobulnicky2004a} above 
which GRBs are suppressed. For the sample with firm redshifts (red line) all critical metallicities 
produce KS $P>0.05$, meaning the data are consistent with no metallicity 
ceiling (note that at $\Zcrit\gtrsim9$ the models are equivalent to 
$\alpha=0$ in Figure \ref{fig:KS}).  However, the peak probability is very similar to the 
metallicity ceiling claimed by \citet{modjaz2008a} 
(dotted line) and the consistency of the model and data rapidly increases for $\Zcrit<8.9$. 
Including the dark GRB sample 
assuming they lie at their maximum possible redshift shifts the $P>0.05$ interval to 
$\Zcrit<8.9$ (formally, this interval equates to a $2-\sigma$ Gaussian equivalent confidence; see
text).  If dark GRBs were all local ($z\approx0$, blue line), no clear value for $\Zcrit$ is 
favored. 
}
\end{figure}

As in \S \ref{section:comparison}, we can formalize this comparison using a one-sample KS test.
Figure \ref{fig:KSZ} shows the KS test probability as a function of $\Zcrit$ in
terms of $12 + \log[\mathrm{O}/\mathrm{H}]$ on the \cite{kobulnicky2004a} scale for 
$8\le\Zcrit\le9.2$.
We find that a cumulative GRB redshift distribution produced by a $\Zcrit\lesssim8.85$ produces 
a redshift-evolution in $\GRBSFR(z)$ that is adequately consistent ($P>0.05$) with the 
firm GRB 
sample (red line), or the sample enlarged by dark GRBs at their redshift limits (orange line).  
Using Equations \ref{eqn:N_GRB} and \ref{eqn:GRBSFR} to construct the likelihood function 
$\mathcal{L}(\Zcrit)$ and the posterior distribution $p(\Zcrit|\bf{z},\GRBSFR)$, we find that
the firm GRB redshift sample is formally $2-\sigma$ consistent with no metallicity ceiling.  
Similarly, we find that including the dark GRB sample at their redshift limits has a 
$2-\sigma$ confidence interval of $\Zcrit<8.9$.
We 
note that in both these models the maximum probability occurs for a metallicity similar to the 
ceiling 
$\Zcrit\approx8.7$ suggested by \citet[][Figure \ref{fig:KSZ} dotted line]{modjaz2008a}.
While the cumulative GRB distribution prefers a $\Zcrit\lesssim8.7$, lower critical metallicities 
display similar consistency.  We note that these low critical metallicities can in principle be 
differentiated based on absolute comparisons of the GRB rate, rather than through
normalized cumulative distribution functions. 
Finally, if dark GRBs are local ($z\sim0$) phenomena 
(blue line), then the 
observed GRB distribution is not very constraining with a large range in $\Zcrit$ displaying 
similar consistency with the data.

In summary, it is relatively straightforward to construct physically-plausible models for $\GRBSFR(z)$ within the sample
uncertainties, based on a metallicity ceiling for GRB production. As the sample sizes grow, there is every prospect of
securing valuable constraints on such models.  As a note, we caution that the recent observations by 
\citet{savaglio2011a} of the afterglow of GRB 090323 at redshift $z\sim4$ shows evidence for two DLAs with supersolar 
metallicities, and that \citet{cenko2011a} suggest at least one of these DLAs may be associated 
with the GRB host galaxy. 
If these observations are confirmed, they could pose difficulty for models of the GRB rate with a low $\Zcrit$.

\begin{figure*}
\figurenum{5}
\epsscale{0.77}
\plotone{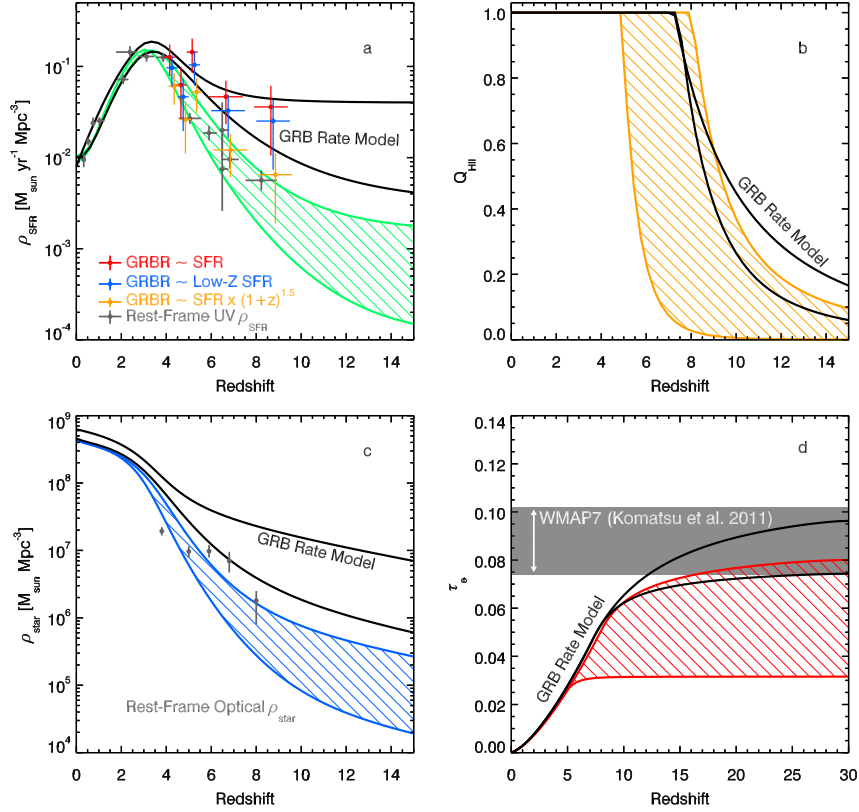}
\caption{\label{fig:sfr}
\scriptsize
Implications of GRB-derived estimates for the high-redshift star formation rate density, $\rhosfr(z)$.  Panel a (upper left) shows
$\rhosfr(z)$ determined from UV-selected galaxies (gray points with error bars, see text) and corresponding parametric SFR 
histories from \citet{robertson2010a} (green hatched region).  Also shown are values implied by the GRB rate assuming 
no evolution in $\GRBSFR(z)$ (red points), our model for GRB production in low-metallicity galaxies (blue points), and 
strong evolution in GRB production per unit star formation rate ($\GRBSFR(z)\propto(1+z)^{1.5}$, orange points). 
The model points have been offset slightly in redshift for clarity and the model error bars reflect Poisson errors on the GRB rate in each redshift bin.  
If the GRB rate to SFR ratio evolves weakly beyond $z>4$ (red and blue points), the rate of discovery of high redshift GRBs already
implies a $\rhosfr(z)$ much larger than that inferred from UV-selected galaxies. Evolution faster than $\GRBSFR(z)\propto(1+z)^{1.5}$ 
would be needed to force agreement. Parameterized star formation histories consistent with the GRB-derived star formation histories in
the constant $\GRBSFR$ and low metallicity star formation models are shown as black lines.  With fiducial choices about the 
character of the stellar populations ($Z\sim0.2Z_{\sun}$), the clumpiness of the intergalactic medium ($C=3$, upper line; 
$C=2.5$ lower line), and the escape fraction of ionizing photons ($\fesc=0.06$, upper line; $\fesc=0.2$, lower line)
we can calculate the reionization history in panel b (upper right) implied by the GRB-derived high-redshift star formation rate 
(black lines) and compare with similar histories calculated by \citet{robertson2010a} determined from UV-selected galaxies 
(orange hatched region).  The GRB-inferred star formation history would produce a large volume filling fraction of 
ionized gas extending to high-redshift.  The path length through this 
ionized gas to the cosmic microwave background provides the optical depth to electron scattering $\tau_{e}$ in panel d (lower right).  
The ionization history computed from the GRB-derived star formation history would easily reach $\tau_{e}$ implied by the seven-year
{\it Wilkinson Microwave Anisotropy Probe} measurements \citep{komatsu2011a}, and produce a much larger value than that
similarly calculated from UV-selected galaxies \citep[red hatched area;][]{robertson2010a}.  While both the ionization
history and the Thomson optical depth depend on specific model choices for $\fesc$ or $C$, the stellar mass density (panel c, 
lower left) is simply determined by the integral of the previous star formation rate density (panel a, upper left).  The stellar mass density to 
$z\sim8$ is shown as gray points with error bars \citep{gonzalez2011a}, with the associated models by \citet[][blue hatched region]{robertson2010a}.  The black lines 
in panel c show the stellar mass density implied by parameterizations of the GRB-derived star formation rate, which clearly 
exceed the stellar mass density at all redshifts.
}
\end{figure*}

\section{The High Redshift Star Formation Rate Density Derived from Distant GRBs}
\label{section:implications}

An exciting development in the past few years has been the discovery and verification of the first
sample of long-duration GRBs beyond redshifts $z\simeq$6 \citep[e.g.,][]{kawai2006a,tanvir2009a,salvaterra2009a,cucchiara2011a}. 
We now utilize our understanding of the ratio of
the GRB rate to the star formation rate density $\GRBSFR$ and its possible redshift dependence to interpret
these data. Of particular interest is how such GRB-derived estimates of the star formation rate density $\rhosfr(z)$
compare with those determined from UV-selected galaxy samples, as these quantities hold implications for
the timing of cosmic reionization and whether the density of star-forming galaxies alone provides sufficient energetic
radiation to reionize the IGM. In the
following we will follow closely the procedures and arguments developed in \citet{robertson2010a}.

In the manner of \cite{yuksel2008a}, we can estimate the star formation rate density as
\begin{equation} 
\label{eqn:sfr_estimate}
\langle \rhosfr \rangle(z_{1},z_{2}) = \frac{\Ngrb(z_{1},z_{2})}{\Ngrb(1,4)} \frac{ \int_{1}^{4} \rhosfr(z) \GRBSFR(z) \frac{\mathrm{d}V}{\mathrm{d}z}\frac{\mathrm{d}z}{1+z} }{ \int_{z_{1}}^{z_{2}} \GRBSFR(z) \frac{\mathrm{d}V}{\mathrm{d}z}\frac{\mathrm{d}z}{1+z} }, 
\end{equation} 
\noindent
where $\Ngrb(z_{1},z_{2})$ is the observed number of GRBs at redshifts $z_{1}<z<z_{2}$.
Panel a of Figure \ref{fig:sfr} shows the resulting comparison.  The star formation rate density determined
from UV-selected galaxies \citep{mclure2010a,bouwens2007a,bouwens2010a,oesch2010a,schiminovich2005a,reddy2009a}
increases to $z\sim3$ and then declines to high redshift \citep[][gray points with error bars]{bouwens2010c}.  Two
parameterized star formation histories from \citet{robertson2010a} consistent with the data
are shown for illustration (green hatched area).  Star formation rate densities estimated from the
high-redshift ($z>4$) GRB rate, as calculated by Equation \ref{eqn:sfr_estimate} and three
models from Figures \ref{fig:Nz} and \ref{fig:Nz.Z} are shown as red ($\alpha=0$, equivalent to $\Zcrit\sim9$), 
blue ($\Zcrit\approx8.7$, equivalent to $\alpha\approx0.5$ at $z\lesssim4$ and $\alpha\approx0$ at $z\gtrsim4$), and 
orange points ($\GRBSFR(z)\propto(1+z)^{1.5}$, stronger evolution than any metallicity evolution model studied in \S
\ref{section:kocevski}). All models have Poisson error bars indicated.  The four lower-$z$ points at $z=4.25$, $4.75$, $5.25$, and $6.75$
contain $N=7$, $3$, $6$, and $4$ GRBs, respectively.  The highest redshift bin ($z\approx8.75$) contains the
two highest-redshift GRBs observed ($z=8.2$, \citealt{tanvir2009a,salvaterra2009a}; $z=9.4$, \citealt{cucchiara2011a}).

Clearly the star formation rate densities estimated from the high redshift GRB rate 
through Equation \ref{eqn:sfr_estimate} for physically-plausible models are
considerably higher than those inferred from UV-selected galaxies.
The model GRB rate to SFR ratio calculated in \S \ref{section:model} has a redshift-dependence that is constant at 
$\GRBSFRl\sim1$ above $z\gtrsim4$, and the 
GRB-inferred $\rhosfr(z)$  at high redshift is therefore large in this model. By considering the recently discovered GRBs at the highest
redshifts \citep{tanvir2009a,salvaterra2009a,cucchiara2011a}, we have extended this result to $z\sim9.5$.  The results are
in contrast to recent claims for a low abundance of $z\gtrsim10$ star-forming galaxies \citep[e.g.,][]{bouwens2011a,oesch2011a}.  
To reconcile $\rhosfr(z)$ estimates from both GRBs and UV-selected galaxies would require
a dramatic evolution in $\GRBSFR(z)$. The physical basis for such an evolution is unclear (c.f., \S \ref{section:model}).

\section{Discussion: Implication for Cosmic Reionization}
\label{section:reionization}

As the discovery of high-redshift galaxies \citep{mclure2010a,bouwens2010a,oesch2010a} and 
quasars \citep{mortlock2011a} reaches beyond $z>7$, it is important to understand the potential role for
star-forming galaxies in reionizaton \citep[for a review, see][]{robertson2010a}.  For instance, \citet{robertson2010a}
have calculated the reionization history, Thomson scattering optical depth, and stellar mass build-up for the 
star formation histories plotted in Figure \ref{fig:sfr} (hatched regions).    The star formation histories in these
models have been parameterized using the formula
\begin{equation}
\label{equation:rhosfr}
\rhosfr(z) = \frac{ a + b\left(z/c\right)^{f}}{1+ \left(z/c\right)^{d}} + g,
\end{equation}
which is the \citet{robertson2010a} generalization of the model by \citet{cole2001a} to include a floor in the
star formation rate.  In Figure \ref{fig:sfr}, panel a, the upper star formation 
history of the green hatched area has parameter values $a=0.009 \Msun\,\,\mathrm{yr}^{-1}\,\,\mathrm{Mpc}^{-3}$, 
$b= 0.27 \Msun\,\,\mathrm{yr}^{-1}\,\,\mathrm{Mpc}^{-3}$, $c=3.7$, $d=7.4$, and 
$g=10^{-3} \Msun\,\,\mathrm{yr}^{-1}\,\,\mathrm{Mpc}^{-3}$.  The lower star formation history model has $c=3.4$,
$d=8.3$, and $g=10^{-4} \Msun\,\,\mathrm{yr}^{-1}\,\,\mathrm{Mpc}^{-3}$.
With metal poor stellar populations, a typical
escape fraction of $\fesc\sim0.3$ and a clumping factor $C\sim2$, the upper $\rhosfr(z)$ curve fully ionizes the intergalactic
medium by $z\sim7$ and recovers the {\it Wilkinson Microwave Anisotropy Probe} (WMAP)
electron scattering optical measurement \citep[e.g.,][]{komatsu2011a}.

Using the same form of Equation \ref{equation:rhosfr} to parameterize the star formation rate density implied 
by the high-redshift GRB rate and a weak-to-moderate redshift-dependence of $\GRBSFR(z)$ 
(Figure \ref{fig:sfr}, panel a, red and blue points), we find that parameter values of 
$a=0.007\Msun\,\,\mathrm{yr}^{-1}\,\,\mathrm{Mpc}^{-3}$,
$b= 0.27 \Msun\,\,\mathrm{yr}^{-1}\,\,\mathrm{Mpc}^{-3}$, $c=3.7$, $d=6.4$, $f=2.5$ and
$g=3\times10^{-3} \Msun\,\,\mathrm{yr}^{-1}\,\,\mathrm{Mpc}^{-3}$ are representative as a ``low-$\rhosfr(z)$''
GRB-derived model (lower black curve), while adopting $d=7.4$ and 
$g=[(4\times10^{-2} - 10^{-3})\times(z/3) + 10^{-3}]\Msun\,\,\mathrm{yr}^{-1}\,\,\mathrm{Mpc}^{-3}$ 
provides representative ``high-$\rhosfr(z)$'' GRB-derived model (upper black curve).

With the same assumptions for the escape fraction, IGM clumping factor, and stellar population model used by 
\citet{robertson2010a}, the star formation history implied by the GRB rate for either a constant $\GRBSFR(z)$ 
or an evolution of the GRB rate to SFR ratio $\GRBSFR(z)$ that tracks star formation in low-metallicity galaxies 
would fully reionize the universe by $z\sim14$ and significantly over-predict the Thomson scattering optical depth 
(e.g., $\tau\sim0.2$). If instead high-redshift galaxies have metallicities $Z\sim0.2Z_{\sun}$ with a \citet{schaerer2003a} stellar population, 
the escape fraction is lower ($\fesc\sim0.06$), and the clumping factor is $C\sim3$, the high-$\rhosfr(z)$ model implied by the 
GRB rate induces reionization by $z\sim7$ and produces a Thomson scattering optical depth which can match the WMAP value 
(upper black curve in Figure \ref{fig:sfr}, panels b and d).  For $\fesc=0.2$ and $C=3$, the low-$\rhosfr(z)$ GRB-derived 
model produces some what lower ionized gas volume filling factors and Thomson optical depths (lower black curve in Figure \ref{fig:sfr}).
Models by \citet{wyithe2010a} reach similar conclusions.

Although the uncertainties associated with calculating $\rhosfr(z)$ from GRBs are large, an explanation of the possible discrepancy
between the GRB-inferred $\rhosfr(z)$ and the abundance of high-redshift galaxies is warranted.
The higher values deduced for the GRB-inferred $\rhosfr(z)$ compared to those inferred from UV-selected galaxies 
may strengthen the case for a steep luminosity
function for the latter and hence closure on the hypothesis that intrinsically faint star-forming galaxies over 6$<z<$12 were
responsible for cosmic reionization \citep[e.g.,][]{yuksel2008a,kistler2009a}.
However, as emphasized by
\citet{robertson2010a}, there is a further constraint provided by the observed stellar mass density at $z\simeq$5-6. This constraint is
determined from rest-frame optical fluxes deduced from the Infrared Array Camera (IRAC) instrument onboard {\it Spitzer Space Telescope}. Importantly, 
the $\rhosfr(z)$ implied by the high redshift GRB rate appears unphysical in that it {\it overproduces the observed stellar mass density 
at $z\gtrsim5$}.  Panel c of Figure \ref{fig:sfr} shows the rest-frame optical stellar mass density determinations by 
\citet[][gray points with error bars; see also \citealt{eyles2005a}, \citealt{stark2007a,stark2009a}, \citealt{labbe2006a,labbe2010a} and \citealt{gonzalez2010a}]{gonzalez2011a}. 
These data may be compared with the integral of the parameterized star formation rate densities consistent by the high-redshift 
GRB-derived $\rhosfr(z)$ (Figure \ref{fig:sfr}, panel c, black lines).  While the escape fraction, clumping factor, and stellar population 
properties are not well known and may be sensibly varied within the broad uncertainties to match the observed Thomson 
optical depth to electron scattering, simultaneously accounting for the comparably low observed stellar mass density at high 
redshift is difficult.  While deeper IRAC observations and other techniques are improving these constraints (Gonzalez et al., in preparation)
it is important to recognize that these new estimates have confirmed that previous efforts have properly accounted for incompleteness 
\citep[e.g.,][]{stark2007a} and the stellar mass density measures are unlikely to increase substantially owing to future data. 

We thus conclude there is an important conflict between fairly reasonable assumptions about how the GRB rate maps to
cosmic star formation and what we already understand about early star formation from UV-selected galaxies. Can the two
probes of early star formation be reconciled? Dust is an unlikely explanation given we would require all UV-selected star 
forming galaxies to be heavily extincted and most studies have utilized their UV continuum slopes to infer the
absence of any significant reddening \citep{bouwens2010b,finkelstein2010a,mclure2011a,dunlop2011a}. 
The high-redshift UV galaxy luminosity functions used to determine the star formation rate density data in Figure \ref{fig:sfr}
are taken from \citet{bouwens2010c}.  At $z=8$ the \citet{bouwens2010c} luminosity function has a measured faint end slope of
$\gamma=-2$, and the luminosity density at $z=8$ shown in Figure \ref{fig:sfr} corresponds to a limiting absolute magnitude of
$M_{\mathrm{AB}}=-18$.  It has been suggested that extending the search to much fainter sources will help bridge the gap between
the galaxy- and GRB-inferred $\rhosfr(z)$ 
(see Figure 4 of \citealt{kistler2009a} and \citealt{choi2011a}), but even with the very steep $\gamma=-2$ the luminosity function
would need to continue down to $M_{\mathrm{AB}}=-(7-9)$ -- slightly brighter than the globular cluster scale -- 
for the UV-luminosity density to recover the GRB-inferred star formation rate
density.
Deeper {\it Hubble Space Telescope} (HST)
data will clarify the possibility. Adjusting the early stellar initial mass function (IMF) so as to increase the luminosity output 
per unit $\rhosfr(z)$ would reduce the stellar mass density; for a \citet{salpeter1955a} IMF with an upper
mass limit of $M_{\mathrm{max}}=100\Msun$, the minimum mass of forming stars would have to increase from $M_{\mathrm{min}}\approx0.1\Msun$
to $M_{\mathrm{min}}\approx2.75-4\Msun$ to decrease the long-lived stellar mass by the required factor of $f_{\star}\sim4-5$. 
Conceivably, the IRAC fluxes at $z\simeq$5-6 are significantly contaminated by
nebular emission so that the deduced stellar masses are over-estimated. An adjustment to the IMF could 
provide as much as $f_{\star}^{2}$ times as much nebular emission contamination in the IRAC bands 
compared to the nebular emission expected from a standard \citet{salpeter1955a} stellar population.
In this case, one factor of $f_{\star}$ increase in the nebular emission contribution to
the IRAC flux would arise from the increase
in the number of Lyman continuum photons produced per unit star formation,
and another factor of $f_{\star}$ would arise from the
decrease in the contribution of long-lived stars to the rest-frame optical emission relative to a 
\citet{salpeter1955a} IMF.
Luminosity evolution in the GRB population could also contribute by altering $\GRBSFR(z)$ 
\citep[e.g.,][]{lloyd-ronning2002a,firmani2004a,kocevski2006a,salvaterra2007a,salvaterra2009b,virgili2011a}, but this
is also uncertain \citep[e.g.,][]{butler2010a,wanderman2010a}.
More likely, there are additional physical
factors affecting the high redshift GRB production rate, i.e.~beyond those of star formation and 
simple metallicity dependence considered in this paper. 
Possible examples include the physics of angular momentum retention in GRB progenitors 
\citep[e.g.,][]{macfadyen2001a} or the effects of the cosmic background radiation temperature on the
initial mass function \citep[e.g.,][]{larson1998a,larson2005a}.
Regardless of what the cause might be, it is clear that the continued discovery and 
study of $z>$6 GRBs promises to shed
light on the reionization process and the assumptions made in the interpretation of star 
formation rates and stellar
masses from UV-selected galaxies.

\section{Summary}
\label{section:summary}

Using the second  {\it Swift} BAT catalog of GRBs \citep{sakamoto2011a}, the observations of dark GRBs 
by \citet{perley2009a}, \citet{greiner2011a}, and \citet{kruhler2011a}, and the
GRB catalogs of \citet{butler2007a,butler2010a}, we have constructed the cumulative redshift distribution $N(<z|\zlim)$ of 
112 luminous ($\Liso>10^{51}$ ergs s$^{-1}$) GRBs out to redshift $z\sim9$.
Comparing with models of an evolving GRB rate to SFR ratio $\GRBSFR(z)$, we find that $N(<z|\zlim)$ constrains
redshift evolution of the form $\GRBSFR(z)\propto(1+z)^{\alpha}$ to $0\lesssim\alpha\lesssim1.5$ out to $\zlim\approx4$.  
By extending the model
of \cite{kocevski2009a} to calculate $\GRBSFR(z)$ from the evolution of the mass-metallicity relation \citep{savaglio2005a}, the star
formation-mass relation \citep{drory2008a} and the stellar mass function \citep{drory2005a}, we find that the presence of a host
galaxy metallicity ceiling $\Zcrit$ above which GRBs are suppressed is highly consistent with the available data if $\Zcrit\lesssim8.85$.  The peak probability of consistency between the model and 
data occurs at  
$\Zcrit\sim8.7$, near the GRB metallicity ceiling suggested by \citet{modjaz2008a}.  Using the method of \citet{yuksel2008a} and 
\citet{kistler2009a}, we use the GRB rate 
at $1<z<4$ (where the star formation rate density $\rhosfr(z)$ is roughly constant) to 
estimate the $\rhosfr(z)$ at $z>4$ 
\citep[including constraints from the highest redshift GRBs, e.g.,][]{kawai2006a,tanvir2009a,salvaterra2009a,cucchiara2011a}.  
We find that for constant to moderate ($\alpha<1$) redshift evolution in $\GRBSFR(z)$, the
star formation rate density predicted by the observed high-redshift GRB rate is substantially higher than the observed $\rhosfr(z)$
inferred from the abundance of UV-selected galaxies \citep[e.g.,][]{bouwens2010a,mclure2010a}, and would over-produce the
observed high-redshift stellar mass density \citep[e.g.,][]{stark2009a,gonzalez2011a}.  Rough agreement between the 
UV- and GRB-determined $\rhosfr(z)$ can be achieved if the redshift-dependence of $\GRBSFR(z)$ is as strong as $\alpha\gtrsim1.5$ 
at $z\gtrsim4$.

\acknowledgments

We thank the anonymous referee for useful suggestions that improved the manuscript.
We also thank Maryam Modjaz and Ehud Nakar for helpful comments and discussions.
BER was supported by a Hubble Fellowship grant, 
program number HST-HF-51262.01-A provided by 
NASA from the Space Telescope Science Institute, 
which is operated by the Association of Universities 
for Research in Astronomy, Incorporated, under NASA 
contract NAS5-26555.

\appendix

Our analysis relies heavily on the excellent GRB catalogs provided by \citet{butler2007a}, \citet{butler2010a}, and the Second {\it Swift} Burst Alert Telescope
catalog \citep{sakamoto2011a}. Further, we make use of the important studies by \citet{perley2009a}, \citet{greiner2011a}, and 
\citet{kruhler2011a} that provided redshift constraints for a sample of dark GRBs.
Below in Table \ref{table:grbs}, we provide the union of these catalogs (162 GRBs in total) as a convenience to the reader only.  The papers providing these catalogs
should be consulted for important
details on the fluence measurements and redshift determinations, along with further original works discovering and characterizing individual GRBs.
The table presents redshifts, isotropic equivalent energies and luminosities, and burst durations as described in \S \ref{section:observed_grbs}.  Of these quantities,
only the isotropic-equivalent energies for 29 GRBs (noted below) and the luminosities (for all GRBs) are newly computed in this work.  Most of the $\Eiso$ values
reported (129) are taken from \citet{butler2007a,butler2010a}, with 4 recent $\Eiso$ values adopted from \citet[][noted below]{sakamoto2011a}.  Of the 29 GRBs whose energies 
we compute as $\Eiso = 4\pi d_{L}^{2} S$, where $d_{L}$ is the luminosity distance to a redshift $z$ and $S$ is the fluence, 21 broad-band fluences are taken from 
\citet{butler2007a,butler2010a} and 8 $10-150$keV fluences (effectively lower limits) are taken from \citet{sakamoto2011a}.  These GRBs are also noted in the table.
Dark GRB redshifts (indicated as upper
limits where appropriate and noted) are adopted from \citet{perley2009a}, \citet{greiner2011a}, or \citet{kruhler2011a}.
Their reported energies and luminosities are calculated assuming the GRBs lie at the redshift
upper limits.  Burst durations mostly are taken from Table 1 of \citet{sakamoto2011a}, except for 
3 values taken from \citet[][noted in table]{butler2007a,butler2010a}.  For completeness reasons as our analysis makes use of 
luminous ($\Liso>10^{51}$ ergs s$^{-1}$) GRBs (see \S \ref{section:observed_grbs}), and less luminous GRBs are therefore indicated with italicized names.

\def\arraystrech{1.5}
\begin{longtable}{lcccc|lcccc}
\tablecolumns{10}
\tablewidth{0pc}
\tablecaption{GRB Catalog}
\tablehead{
\colhead{GRB} & \colhead{z} & \colhead{$\Eiso$ [$10^{52}$ ergs]} & \colhead{$t_{90}$ [s]} &\colhead{$\Liso$ [$10^{52}$ ergs s$^{-1}$]}&\colhead{GRB} & \colhead{z} & \colhead{$\Eiso$ [$10^{52}$ ergs]} & \colhead{$t_{90}$ [s]} &\colhead{$\Liso$ [$10^{52}$ ergs s$^{-1}$]}
\label{table:grbs}}
\tabletypesize{\scriptsize}
\startdata
{\it 050126}& 1.29&$ 0.80_{-0.20 }^{+1.00}$	& 48.0 	&$ 0.038_{-0.010}^{+0.048}$	  	&     070521	& 1.35\tnm{c}&$ 25.2_{-8.8}^{+22.0}$\tnm{a}& 38.6 	&$ 1.54_{-0.53 }^{+1.34 }$\\
{\it 050223}	& 0.58		&$ 0.070_{-0.010 }^{+0.050 }$	& 21.7 	&$ 0.0051_{-0.0007 }^{+0.0037}$		&     070529	& 2.50	&$ 9.0_{-3.0 }^{+9.0 }$		& 109 	&$ 0.29_{-0.10 }^{+0.29 }$\\
     050315 	& 1.95		&$ 5.7_{-0.1 }^{+6.2 }$		& 95.6 	&$ 0.18_{-0.003}^{+0.19}$		&     070611	& 2.04	&$ 0.50_{-0.10 }^{+0.40 }$	& 13.2 	&$ 0.12_{-0.02 }^{+0.09 }$\\
{\it 050318}	& 1.44		&$ 1.2_{-0.2 }^{+0.2 }$		& 40.0 	&$ 0.073_{-0.012 }^{+0.012 }$		&{\it 070612A}	& 0.62	&$ 2.0_{-0.4 }^{+1.8 }$		& 365 	&$ 0.0089_{-0.0018 }^{+0.0080}$\\
     050319	& 3.24		&$ 4.6_{-0.6 }^{+6.5 }$		& 152 	&$ 0.13_{-0.02 }^{+0.18 }$		&     070721B	& 3.63	&$ 30.0_{-10.0 }^{+20.0 }$	& 337 	&$ 0.41_{-0.14 }^{+0.27 }$\\
     050401	& 2.90		&$ 32.0_{-7.0 }^{+26.0 }$	& 33.3 	&$ 3.75_{-0.82 }^{+3.04 }$		&     070802	& 2.45	&$ 0.50_{-0.10 }^{+0.52 }$	& 15.8 	&$ 0.11_{-0.02 }^{+0.11 }$\\
     050412	&<4.50\tnm{b} 	&$ 101_{-61}^{+183}$\tnm{a}	& 26.5 	&$ 21.1_{-12.7}^{+38.0}$		&     070810A	& 2.17	&$ 0.90_{-0.10 }^{+0.30 }$	& 9.04 	&$ 0.32_{-0.04 }^{+0.11 }$\\
{\it 050416A}	& 0.65		&$ 0.10_{-0.02 }^{+0.05}$	& 6.62 	&$ 0.025_{-0.005}^{+0.012}$		&     071003	& 1.60	&$ 18.0_{-6.0 }^{+14.0 }$	& 148 	&$ 0.32_{-0.11 }^{+0.25 }$\\
     050505	& 4.28		&$ 16.0_{-3.0 }^{+13.0 }$	& 58.9 	&$ 1.43_{-0.27 }^{+1.16 }$		&{\it 071010A}	& 0.98	&$ 0.13_{-0.02 }^{+0.24 }$	& 6.32 	&$ 0.041_{-0.006 }^{+0.075 }$\\
     050525A	& 0.61		&$ 2.0_{-0.1 }^{+0.1 }$		& 8.84 	&$ 0.37_{-0.02 }^{+0.02 }$		&{\it 071010B}	& 0.95	&$ 1.8_{-0.1}^{+0.4}$		& 36.1 	&$ 0.097_{-0.005 }^{+0.022 }$\\
     050603	& 2.82		&$ 50.0_{-20.0 }^{+40.0 }$	& 22.0 	&$ 8.68_{-3.47 }^{+6.95 }$		&     071011	&<5.00\tnm{b}&$ 234_{-104}^{+182}$\tnm{a}& 80.9 	&$ 17.4_{-7.7}^{+13.5}$\\
     050607	&<4.00\tnm{b}	&$ 12.3_{-1.5 }^{+10.8}$\tnm{a}	& 48.0 	&$ 1.28_{-0.16 }^{+1.12 }$		&     071020	& 2.15	&$ 10.0_{-3.0}^{+10.0}$		& 4.30 	&$ 7.32_{-2.20 }^{+7.32 }$\\
     050713A	&<3.60\tnm{b}	&$ 156_{-48}^{+132}$\tnm{a}	& 94.9 	&$ 7.54_{-2.32 }^{+6.38 }$		&     071025	& 5.20	&$ 428_{-86}^{+285}$\tnm{a}	& 241 	&$ 11.0_{-2.2 }^{+7.3 }$\\
     050730	& 3.97		&$ 9.0_{-3.0 }^{+8.0 }$		& 145 	&$ 0.31_{-0.10 }^{+0.27 }$		&{\it 071031}	& 2.69	&$ 3.9_{-0.6}^{+4.1}$		& 150 	&$ 0.096_{-0.015 }^{+0.101 }$\\
{\it 050801}	& 1.38		&$ 0.41_{-0.06 }^{+0.64 }$\tnm{a}& 19.4	&$ 0.050_{-0.007 }^{+0.078 }$		&     071117	& 1.33	&$ 1.9_{-0.3}^{+0.8}$		& 6.07 	&$ 0.73_{-0.12 }^{+0.31 }$\\
     050802	& 1.71		&$ 1.8_{-0.3 }^{+1.6 }$		& 27.5 	&$ 0.18_{-0.03 }^{+0.16 }$		&{\it 071122}	& 1.14	&$ 0.30_{-0.10 }^{+0.50 }$	& 80.0 	&$ 0.0080_{-0.0027 }^{+0.0134 }$\\
{\it 050803}	& 0.42		&$ 0.24_{-0.08 }^{+0.24 }$	& 88.1 	&$ 0.0039_{-0.0013}^{+0.0039}$	&     080129	& 4.35	&$ 8.0_{-4.0 }^{+7.0 }$		& 50.2 	&$ 0.85_{-0.43 }^{+0.75 }$\\
     050814	& 5.30		&$ 6.0_{-1.0 }^{+3.0 }$		& 144 	&$ 0.26_{-0.04 }^{+0.13 }$		&     080210	& 2.64	&$ 5.1_{-0.9 }^{+4.5 }$		& 39.4 	&$ 0.47_{-0.08 }^{+0.42 }$\\
     050820A	& 2.61		&$ 20.0_{-10.0 }^{+20.0 }$	& 240\tnm{f} 	&$ 0.30_{-0.15 }^{+0.30 }$	&{\it 080310}	& 2.43	&$ 5.9_{-1.0 }^{+10.5 }$	& 352 	&$ 0.057_{-0.010 }^{+0.102 }$\\
{\it 050824}	& 0.83		&$ 0.15_{-0.04 }^{+0.86 }$	& 24.8 	&$ 0.011_{-0.003 }^{+0.063 }$		&     080319A	&<2.20\tnm{b}	&$ 29.2_{-3.7}^{+25.5}$\tnm{a}& 43.6 	&$ 2.14_{-0.27 }^{+1.87 }$\\
{\it 050826}	& 0.30		&$ 0.03_{-0.02 }^{+0.04 }$	& 35.7 	&$ 0.0011_{-0.0007}^{+0.0015}$	&     080319B	& 0.94	&$ 400_{-100}^{+200}$		& 125 	&$ 6.21_{-1.55 }^{+3.10 }$\\
     050904	& 6.29		&$ 130_{-40}^{+70}$		& 182 	&$ 5.22_{-1.61 }^{+2.81 }$		&     080319C	& 1.95	&$ 6.0_{-1.0 }^{+5.0 }$		& 29.5 	&$ 0.60_{-0.10 }^{+0.50 }$\\
     050908	& 3.35		&$ 1.3_{-0.3 }^{+0.9 }$		& 18.3 	&$ 0.31_{-0.07 }^{+0.21 }$		&     080320 	&<7.00\tnm{b}&$ 34.2_{-5.7}^{+45.6}$\tnm{a}& 13.8&$ 19.8_{-3.3 }^{+26.4 }$\\
{\it 050915A}	& 0.40\tnm{c}	&$ 0.073_{-0.023}^{+0.102}$\tnm{a}&53.4	&$ 0.0019_{-0.0006}^{+0.0027}$	&{\it 080330}	& 1.51	&$ 0.41_{-0.06 }^{+0.94 }$	& 67.1 	&$ 0.015_{-0.002 }^{+0.035 }$\\
     050922C	& 2.20		&$ 3.9_{-0.8 }^{+2.7 }$		& 4.54 	&$ 2.75_{-0.56 }^{+1.90 }$		&     080411	& 1.03	&$ 23.0_{-4.0 }^{+0.9 }$	& 56.3 	&$ 0.83_{-0.14 }^{+0.03 }$\\
{\it 051016B}	& 0.94		&$ 0.037_{-0.006 }^{+0.056 }$	& 4.02 	&$ 0.018_{-0.003 }^{+0.027 }$		&     080413A	& 2.43	&$ 9.0_{-2.0 }^{+6.0 }$	& 46.4 	&$ 0.67_{-0.15 }^{+0.44 }$\\
     051109A	& 2.35 		&$ 2.3_{-0.5 }^{+2.4 }$		& 37.2 	&$ 0.21_{-0.04 }^{+0.22 }$		&     080413B	& 1.10	&$ 1.5_{-0.2 }^{+0.2 }$	& 8.00 	&$ 0.39_{-0.05 }^{+0.05 }$\\
{\it 051109B}	& 0.08		&$ 0.00036_{-0.00009}^{+0.00019}$& 13.4 &$ 0.00003_{-0.00001 }^{+0.00002 }$	&{\it 080430}	& 0.77	&$ 0.38_{-0.08 }^{+0.30 }$	& 14.2 	&$ 0.047_{-0.010 }^{+0.037 }$\\
     051111	& 1.55		&$ 6.0_{-2.0 }^{+5.0 }$		& 64.0 	&$ 0.24_{-0.08 }^{+0.20 }$		&     080516	& 3.60\tnm{g}	&$ 12.0_{-4.8}^{+6.0}$\tnm{a}	& 5.75 	&$ 9.57_{-3.83 }^{+4.79 }$\\
     060108	& 2.03		&$ 0.59_{-0.08 }^{+0.84 }$	& 14.2 	&$ 0.13_{-0.02 }^{+0.18 }$		&{\it 080520}	& 1.55	&$ 0.11_{-0.04}^{+1.45}$	& 3.32 	&$ 0.084_{-0.031 }^{+1.112 }$\\
     060110	&<5.00\tnm{b}	&$ 83.3_{-15.6}^{+67.7}$\tnm{a}	& 21.1 &$ 23.6_{-4.4}^{+19.2}$			&     080603B	& 2.69	&$ 6.0_{-1.0}^{+1.0}$	& 59.1 	&$ 0.37_{-0.06 }^{+0.06 }$\\
     060115	& 3.53		&$ 6.0_{-1.0}^{+2.0}$		& 122 	&$ 0.22_{-0.04 }^{+0.07 }$		&{\it 080604}	& 1.42	&$ 0.70_{-0.10 }^{+0.80 }$	& 69.2 	&$ 0.024_{-0.003 }^{+0.028 }$\\
     060116	& 6.60		&$ 21.0_{-7.0}^{+16.0}$		& 105 	&$ 1.52_{-0.51 }^{+1.16 }$		&     080605	& 1.64	&$ 21.0_{-4.0}^{+9.0}$	& 18.0 	&$ 3.07_{-0.59 }^{+1.32 }$\\
{\it 060124}	& 2.30		&$ 0.70_{-0.10}^{+0.70}$	& 658 	&$ 0.0035_{-0.0005}^{+0.0035}$	&     080607	& 3.04	&$ 280_{-90}^{+130}$	& 78.9 	&$ 14.3_{-4.6}^{+6.7}$\\
{\it 060202}	& 0.78		&$ 0.70_{-0.10}^{+0.60 }$	& 172 	&$ 0.0073_{-0.0010}^{+0.0062}$	&{\it 080707}	& 1.23	&$ 0.34_{-0.05 }^{+0.41 }$	& 30.2 	&$ 0.025_{-0.004 }^{+0.030 }$\\
     060206	& 4.06		&$ 4.1_{-0.7}^{+1.2}$		& 7.55 	&$ 2.75_{-0.47 }^{+0.80 }$		&{\it 080710}	& 0.85	&$ 0.80_{-0.40 }^{+0.80 }$	& 112 	&$ 0.013_{-0.007 }^{+0.013 }$\\
     060210	& 3.91		&$ 42.0_{-8.0 }^{+35.0 }$	& 242 	&$ 0.85_{-0.16 }^{+0.71 }$		&     080721	& 2.59	&$ 110_{-50}^{+110}$	& 176 	&$ 2.24_{-1.02 }^{+2.24 }$\\
{\it 060218}	& 0.03		&$ 0.00029_{-0.00007}^{+0.00014}$& 128\tnm{f} 	&$ 2.3\times10^{-6}$		&     080804	& 2.20	&$ 16.0_{-7.0}^{+17.0}$	& 37.2 	&$ 1.38_{-0.60 }^{+1.46 }$\\
     060223A	& 4.41		&$ 3.1_{-0.5 }^{+1.2 }$		& 11.3 	&$ 1.48_{-0.24 }^{+0.57 }$		&{\it 080805}	& 1.50	&$ 4.0_{-2.0 }^{+2.0 }$	& 107 	&$ 0.094_{-0.047 }^{+0.047 }$\\
     060418	& 1.49		&$ 10.0_{-2.0 }^{+7.0 }$	& 109 	&$ 0.23_{-0.05 }^{+0.16 }$		&     080810	& 3.36	&$ 30.0_{-20.0 }^{+20.0 }$	& 108 	&$ 1.21_{-0.81 }^{+0.81 }$\\
{\it 060428B}	& 0.35		&$ 0.020_{-0.004 }^{+0.019 }$	& 96.0 	&$ 0.00028_{-0.00006 }^{+0.00027 }$	&     080905B	& 2.37	&$ 3.4_{-0.6 }^{+3.1 }$	& 102 	&$ 0.11_{-0.02 }^{+0.10 }$\\
     060502A	& 1.50		&$ 3.2_{-0.9 }^{+2.8 }$		& 28.5 	&$ 0.28_{-0.08 }^{+0.25 }$		&     080913	& 6.70	&$ 7.0_{-1.0 }^{+7.0 }$	& 7.46 	&$ 7.23_{-1.03 }^{+7.23 }$\\
     060510B	& 4.94		&$ 23.0_{-4.0 }^{+10.0 }$	& 263 	&$ 0.52_{-0.09 }^{+0.23 }$		&{\it 080916A}	& 0.69	&$ 0.81_{-0.09 }^{+0.21 }$	& 61.3 	&$ 0.022_{-0.002 }^{+0.006 }$\\
     060512	& 2.10		&$ 0.81_{-0.16}^{+0.16}$\tnm{e}	& 11.4 	&$ 0.22_{-0.04 }^{+0.04 }$		&{\it 080928}	& 1.69	&$ 2.8_{-0.5}^{+2.4}$	& 234 	&$ 0.032_{-0.006 }^{+0.028 }$\\
     060522	& 5.11		&$ 7.0_{-1.0 }^{+7.0 }$		& 69.1 	&$ 0.62_{-0.09 }^{+0.62 }$		&{\it 081007}	& 0.53	&$ 0.07_{-0.01}^{+0.05}$	& 9.01 	&$ 0.012_{-0.002 }^{+0.008 }$\\
{\it 060526}	& 3.22		&$ 5.2_{-0.4 }^{+5.6 }$		& 275 	&$ 0.080_{-0.006 }^{+0.086 }$		&{\it 081008}	& 1.97	&$ 6.0_{-1.0 }^{+3.0 }$	& 180 	&$ 0.099_{-0.017 }^{+0.050 }$\\
{\it 060604}	& 2.68		&$ 0.50_{-0.10 }^{+1.20 }$	& 96.0 	&$ 0.019_{-0.004 }^{+0.046 }$		&     081028	& 3.04	&$ 11.0_{-2.0}^{+3.0}$	& 284 	&$ 0.16_{-0.03 }^{+0.04 }$\\
{\it 060605}	& 3.77		&$ 2.5_{-0.6 }^{+3.1 }$		& 539 	&$ 0.022_{-0.005 }^{+0.027 }$		&     081029	& 3.85	&$ 15.0_{-7.0}^{+9.0}$	& 275 	&$ 0.26_{-0.12 }^{+0.16 }$\\
     060607A	& 3.07		&$ 9.0_{-2.0 }^{+7.0 }$		& 103 	&$ 0.36_{-0.08 }^{+0.28 }$		&{\it 081109}	& 0.98\tnm{h}	&$ 4.1_{-2.2}^{+2.6}$\tnm{a}	& 221 	&$ 0.037_{-0.020 }^{+0.023 }$\\
{\it 060614}	& 0.13		&$ 0.24_{-0.04 }^{+0.04 }$	& 109 	&$ 0.0025_{-0.0004 }^{+0.0004}$		&     081118	& 2.58	&$ 2.8_{-0.4 }^{+4.4 }$	& 49.3 	&$ 0.20_{-0.03 }^{+0.32 }$\\
     060707	& 3.42		&$ 6.1_{-0.9 }^{+1.9 }$		& 66.7 	&$ 0.40_{-0.06 }^{+0.13 }$		&     081121	& 2.51	&$ 16.0_{-4.0 }^{+15.0 }$	& 17.7 	&$ 3.18_{-0.80 }^{+2.98 }$\\
     060708	& 1.92		&$ 0.60_{-0.10 }^{+0.40 }$	& 9.96 	&$ 0.18_{-0.03 }^{+0.12 }$		&     081203A	& 2.05	&$ 17.0_{-4.0 }^{+13.0 }$	& 223 	&$ 0.23_{-0.05 }^{+0.18 }$\\
     060714	& 2.71		&$ 7.7_{-0.9 }^{+7.5 }$		& 116 	&$ 0.25_{-0.03 }^{+0.24 }$		&     081222	& 2.77	&$ 15.0_{-2.0 }^{+3.0 }$	& 33.0 	&$ 1.71_{-0.23 }^{+0.34 }$\\
{\it 060729}	& 0.54		&$ 0.33_{-0.06 }^{+0.29 }$	& 113 	&$ 0.0045_{-0.0008}^{+0.0040}$		&     081228	& 3.40\tnm{g}	&$ 3.7_{-1.3 }^{+1.6 }$\tnm{a}	& 3.00 	&$ 5.36_{-1.84 }^{+2.30 }$\\
     060805A	&<3.80\tnm{b}	&$ 1.8_{-0.5}^{+2.7}$\tnm{a}	& 4.93 &$ 1.72_{-0.53 }^{+2.65 }$		&     090102	& 1.55	&$ 14.0_{-5.0 }^{+10.0 }$	& 29.3 	&$ 1.22_{-0.43 }^{+0.87 }$\\
     060814	& 0.84		&$ 11.6_{-2.0 }^{+5.8 }$\tnm{a}	& 145 	&$ 0.15_{-0.03 }^{+0.07 }$		&     090205	& 4.65	&$ 1.2_{-0.2 }^{+1.6 }$	& 8.80 	&$ 0.77_{-0.13 }^{+1.03 }$\\
{\it 060904B}	& 0.70		&$ 0.30_{-0.06 }^{+0.19 }$	& 172 	&$ 0.0030_{-0.0006}^{+0.0019}$		&     090313	& 3.38	&$ 4.6_{-0.5 }^{+7.0 }$	& 70.7 	&$ 0.28_{-0.03 }^{+0.43 }$\\
     060906	& 3.69		&$ 13.0_{-1.0 }^{+12.0 }$	& 44.6 	&$ 1.37_{-0.11 }^{+1.26 }$		&{\it 090417B}	& 0.34	&$ 0.16_{-0.03 }^{+0.16 }$\tnm{a}	& 282 	&$ 0.00074_{-0.00015 }^{+0.00076 }$\\
     060908	& 1.88		&$ 8.1_{-4.5 }^{+1.9 }$\tnm{d}	& 18.8 	&$ 1.24_{-0.69 }^{+0.29 }$		&     090418A	& 1.61	&$ 9.0_{-3.0 }^{+6.0 }$	& 56.3 	&$ 0.42_{-0.14 }^{+0.28 }$\\
     060912A	& 0.94		&$ 0.80_{-0.20 }^{+0.50 }$	& 5.03 	&$ 0.31_{-0.08 }^{+0.19 }$		&     090423	& 8.23	&$ 8.0_{-1.0 }^{+2.0 }$	& 9.77 	&$ 7.56_{-0.94 }^{+1.89 }$\\
     060923A	&<4.00\tnm{b}	&$ 20.0_{-4.6 }^{+9.2 }$\tnm{a}& 51.5 &$ 1.94_{-0.45 }^{+0.89 }$		&{\it 090424}	& 0.54	&$ 2.6_{-0.4 }^{+0.4 }$	& 49.5 	&$ 0.081_{-0.012 }^{+0.012 }$\\
     060926	& 3.21		&$ 1.0_{-0.2 }^{+2.2 }$		& 7.79 	&$ 0.54_{-0.11 }^{+1.19 }$		&     090429B	& 9.40	&$ 54.9_{-7.8}^{+15.7}$\tnm{a}	& 5.61 	&$ 101.7_{-14.5 }^{+29.1}$\\
     060927	& 5.46		&$ 9.0_{-1.0 }^{+2.0 }$		& 22.4 	&$ 2.59_{-0.29 }^{+0.58 }$		&     090510	& 0.90	&$ 0.30_{-0.20 }^{+0.50 }$	& 5.66 	&$ 0.10_{-0.07 }^{+0.17 }$\\
     061004	& 3.30		&$ 2.0_{-0.3}^{+1.1 }$		& 6.26 	&$ 1.37_{-0.21 }^{+0.76 }$		&     090516	& 4.11	&$ 50.0_{-10.0 }^{+50.0 }$	& 208 	&$ 1.23_{-0.25 }^{+1.23 }$\\
     061007	& 1.26		&$ 140_{-60 }^{+110}$		& 75.7 	&$ 4.18_{-1.79 }^{+3.29 }$		&     090519	& 3.85	&$ 15.0_{-8.0 }^{+13.0 }$	& 58.3 	&$ 1.25_{-0.67 }^{+1.08 }$\\
{\it 061021}	& 0.35		&$ 0.40_{-0.16 }^{+0.32}$\tnm{a}& 43.8 	&$ 0.012_{-0.005 }^{+0.010 }$		&     090529A	& 2.62	&$ 2.5_{-0.5 }^{+1.4 }$		& 80.0\tnm{f} 	&$ 0.11_{-0.02 }^{+0.06 }$\\
{\it 061110A}	& 0.76		&$ 0.28_{-0.06 }^{+0.28}$	& 44.5 	&$ 0.011_{-0.002 }^{+0.011 }$		&     090618	& 0.54	&$ 15.0_{-1.0 }^{+1.0 }$	& 113 	&$ 0.20_{-0.01 }^{+0.01 }$\\
     061110B	& 3.43		&$ 13.0_{-6.0 }^{+16.0}$	& 133 	&$ 0.43_{-0.20 }^{+0.53 }$		&     090715B	& 3.00	&$ 24.0_{-5.0 }^{+15.0 }$	& 265 	&$ 0.36_{-0.08 }^{+0.23 }$\\
     061121	& 1.31		&$ 19.0_{-5.0 }^{+11.0}$	& 81.2 	&$ 0.54_{-0.14 }^{+0.31 }$		&     090726	& 2.71	&$ 1.8_{-0.4 }^{+2.1 }$		& 56.7 	&$ 0.12_{-0.03 }^{+0.14 }$\\
     061126	& 1.16		&$ 8.0_{-2.0}^{+7.0}$		& 50.3 	&$ 0.34_{-0.09 }^{+0.30 }$		&     090809A	& 2.74	&$ 1.4_{-0.4 }^{+2.4 }$		& 7.84 	&$ 0.67_{-0.19 }^{+1.14 }$\\
     061222A	& 2.09		&$ 67.4_{-12.8}^{+35.3}$\tnm{a}	& 96.0 	&$ 2.17_{-0.41 }^{+1.14 }$		&     090812	& 2.45	&$ 19.0_{-5.0 }^{+17.0 }$	& 75.1 	&$ 0.87_{-0.23 }^{+0.78 }$\\
     061222B	& 3.35		&$ 8.0_{-2.0 }^{+7.0 }$		& 37.2 	&$ 0.94_{-0.23 }^{+0.82 }$		&{\it 090814A}	& 0.70	&$ 0.27_{-0.03 }^{+0.03 }$\tnm{e}	& 78.2 	&$ 0.0059_{-0.0007}^{+0.0007}$\\
     070110	& 2.35		&$ 43.0_{-0.5 }^{+2.5 }$	& 79.7 	&$ 1.81_{-0.02 }^{+0.11 }$		&     090904B	&<5.00\tnm{i}	&$ 34.4_{-14.0 }^{+14.0 }$\tnm{e}	& 64.0 	&$ 3.23_{-1.31 }^{+1.31 }$\\
{\it 070208}	& 1.17		&$ 0.28_{-0.08 }^{+0.22 }$	& 64.0 	&$ 0.0095_{-0.0027}^{+0.0074 }$		&     090926B	& 1.24	&$ 5.4_{-2.0 }^{+2.8 }$\tnm{d}	& 99.3 	&$ 0.12_{-0.05 }^{+0.06 }$\\
{\it 070306}	& 1.50		&$ 6.0_{-1.0 }^{+5.0 }$		& 209 	&$ 0.072_{-0.012 }^{+0.060 }$		&     090927	& 1.37	&$ 0.23_{-0.03 }^{+0.03 }$\tnm{e}	& 2.16 	&$ 0.25_{-0.03 }^{+0.03 }$\\
{\it 070318}	& 0.84		&$ 0.90_{-2.00 }^{+0.90 }$	& 132 	&$ 0.013_{-0.028 }^{+0.013 }$		&     091018	& 0.97	&$ 0.70_{-0.10 }^{+0.30 }$\tnm{d}	& 4.37 	&$ 0.32_{-0.05 }^{+0.14 }$\\
     070411	& 2.95		&$ 10.0_{-2.0 }^{+8.0 }$	& 102 	&$ 0.39_{-0.08 }^{+0.31 }$		&     091020	& 1.71	&$ 7.5_{-0.3}^{+0.3}$\tnm{e}	& 38.9 	&$ 0.52_{-0.02 }^{+0.02 }$\\
{\it 070419A}	& 0.97		&$ 0.24_{-0.05 }^{+0.23 }$	& 160 	&$ 0.0030_{-0.0006 }^{+0.0028}$		&{\it 091024}	& 1.09	&$ 3.9_{-0.2}^{+0.2}$\tnm{e}	& 112 	&$ 0.072_{-0.003 }^{+0.003 }$\\
     070506	& 2.31		&$ 0.26_{-0.05 }^{+0.17 }$	& 4.35 	&$ 0.20_{-0.04 }^{+0.13 }$		&     091029	& 2.75	&$ 8.5_{-2.5}^{+4.5}$\tnm{d}	& 39.2 	&$ 0.81_{-0.24 }^{+0.43 }$\\
     070508	& 0.82		&$ 8.0_{-1.0 }^{+2.0 }$		& 20.9 	&$ 0.70_{-0.09 }^{+0.17 }$		&     091127	& 0.49	&$ 0.77_{-0.02 }^{+0.02 }$\tnm{e}	& 7.42 	&$ 0.16_{-0.00 }^{+0.00 }$\\
{\it 070518}	& 1.16		&$ 0.090_{-0.010 }^{+0.150 }$	& 4.35 	&$ 0.045_{-0.005 }^{+0.074 }$		&     091208B	& 1.06	&$ 1.9_{-0.1}^{+0.1}$\tnm{e}	& 14.8 	&$ 0.27_{-0.02 }^{+0.02 }$\\
\end{longtable}
\noindent
$^a$$\Eiso$ calculated from fluence provided by \citet{butler2007a,butler2010a}.\\
$^b$Dark GRB redshift limit from \citet{perley2009a}.\\
$^c$Redshift from \citet{perley2009a}.\\
$^d$$\Eiso$ from Table 13 of \citet{sakamoto2011a}.\\
$^e$$\Eiso$ calculated from fluence provided by Table 2 of \citet{sakamoto2011a}.\\
$^f$Burst duration taken from \citet{butler2007a,butler2010a}.\\
$^g$Redshift from \citet{greiner2011a}.\\
$^h$Redshift from \citet{kruhler2011a}.\\
$^i$Dark GRB redshift limit from \citet{greiner2011a}.
\end{document}